\documentclass[sn-mathphys,Numbered]{sn-jnl}
\usepackage{graphicx}%
\usepackage{multirow}%
\usepackage{amsmath,amssymb,amsfonts}%
\usepackage{amsthm}%
\usepackage{bbm}
\usepackage{mathrsfs}%
\usepackage[title]{appendix}%
\usepackage{xcolor}%
\usepackage{textcomp}%
\usepackage{manyfoot}%
\usepackage{booktabs}%
\usepackage{algpseudocode}%
\usepackage[linesnumbered,ruled,vlined]{algorithm2e}
\usepackage{listings}%
\usepackage{array}
\usepackage{url}
\newcolumntype{Q}{>{\centering$}p{0.4cm}<{$}}

\newtheorem{thm}{Theorem}[section]
\newtheorem{cor}[thm]{Corollary}
\newtheorem{prop}[thm]{Proposition}
\newtheorem{lem}[thm]{Lemma}
\newtheorem{conj}[thm]{Conjecture}

\newtheorem{constr}[thm]{Construction}

\newtheorem{defn}{Definition}

\newcommand{\bit}{\begin{itemize}}
\newcommand{\eit}{\end{itemize}}
\newcommand{\bcor}{\begin{cor}}
\newcommand{\ecor}{\end{cor}}
\newcommand{\beq}{\begin{equation}}
\newcommand{\eeq}{\end{equation}}
\newcommand{\beqn}{\begin{equation}}
\newcommand{\eeqn}{\end{equation}}
\newcommand{\bea}{\begin{eqnarray}}
\newcommand{\eea}{\end{eqnarray}}
\newcommand{\bean}{\begin{eqnarray*}}
\newcommand{\eean}{\end{eqnarray*}}
\newcommand{\ben}{\begin{enumerate}}
\newcommand{\een}{\end{enumerate}}
\newcommand{\bdefn}{\begin{defn}}
\newcommand{\edefn}{\end{defn}}
\newcommand{\bprop}{\begin{prop}}
\newcommand{\eprop}{\end{prop}}
\newcommand{\blem}{\begin{lem}}
\newcommand{\elem}{\end{lem}}
\newcommand{\bthm}{\begin{thm}}
\newcommand{\ethm}{\end{thm}}
\newcommand{\bconj}{\begin{conj}}
\newcommand{\econj}{\end{conj}}
\newcommand{\bpf}{\begin{proof}}
\newcommand{\epf}{\end{proof}}
\newcommand{\bconstr}{\begin{constr}}
\newcommand{\econstr}{\end{constr}}

\newcommand{\gap}{{\sf{gap}}}
\newcommand{\gaps}{{\bf{g}}}

\newcommand{\bx}{{\bf{x}}}
\newcommand{\bxo}{{\bf{x}_1}}
\newcommand{\bxt}{{\bf{x}_2}}

\newcommand{\bxl}{{\bf{x}_{\ell}}}
\newcommand{\bxlmo}{{\bf{x}_{\ell-1}}}
\newcommand{\bxlmt}{{\bf{x}_{\ell-2}}}

\newcommand{\bc}{{\bf{c}}}
\newcommand{\dec}{\textsf{dec}}
\newcommand{\pos}{\textsf{pos}}

\begin{document}
\title{\centering Binary Constant Weight Codes with \\ Low-Complexity Encoding and Decoding}
\author*[1]{\fnm{Birenjith} \sur{Sasidharan}}\email{birenjith.padmakumarisasidharan@monash.edu}

\author[1]{\fnm{Emanuele} \sur{Viterbo}}\email{emanuele.viterbo@monash.edu}

\author[2]{\fnm{Son Hoang} \sur{Dau}}\email{sonhoang.dau@rmit.edu.au}

\affil*[1]{\orgdiv{ECSE Dept.}, \orgname{Monash University}, \orgaddress{
\city{Clayton}, 
\state{Victoria}, \country{Australia}}}

 \affil[2]{
 \orgname{RMIT University}, \orgaddress{
\city{Melbourne}, 
\state{Victoria}, \country{Australia}}}


\abstract{In this paper, we focus on the design of binary constant weight codes that admit low-complexity encoding and decoding algorithms, and that have size $M=2^k$ so that codewords can conveniently be labeled with binary vectors of length $k$. For every integer $\ell \geq 3$, we construct a  $(n=2^\ell, M=2^{k_{\ell}}, d=2)$ constant weight code ${\cal C}[\ell]$ of weight $\ell$ by encoding information in the gaps between successive $1$'s of a vector. The code is associated with a finite integer sequence of length $\ell$ satisfying a constraint defined as {\em anchor-decodability} that is pivotal to ensure low complexity for encoding and decoding. The time complexity of the encoding algorithm is linear in the input size $k$, and that of the decoding algorithm is poly-logarithmic in the input size $n$, discounting the linear time spent on parsing the input. Both the algorithms do not require expensive computation of binomial coefficients, unlike the case in many existing schemes. Among codes generated by all anchor-decodable sequences, we show that ${\cal C}[\ell]$ has the maximum size with $k_{\ell} \geq \ell^2-\ell\log_2\ell + \log_2\ell - 0.279\ell - 0.721$. As $k$ is upper bounded by $\ell^2-\ell\log_2\ell +O(\ell)$ information-theoretically, the code ${\cal C}[\ell]$ is optimal in its size with respect to two higher order terms of $\ell$. In particular, $k_\ell$ meets the upper bound for $\ell=3$ and one-bit away for $\ell=4$. On the other hand, we show that ${\cal C}[\ell]$ is not unique in attaining $k_{\ell}$ by constructing an alternate code ${\cal \hat{C}}[\ell]$ again parameterized by an integer $\ell \geq 3$ with a different low-complexity decoder, yet having the same size $2^{k_{\ell}}$ when $3 \leq \ell \leq 7$. Finally, we also derive new codes by modifying ${\cal C}[\ell]$ that offer a wider range on blocklength and weight while retaining low complexity for encoding and decoding. For certain selected values of parameters, these modified codes too have an optimal $k$.}



\maketitle

\section{Introduction}

Let 
$n$ and $w\leq n$ be positive integers. A constant weight binary $(n,M,d)$ code ${\cal C}$ of blocklength $n$ and weight $w$ is defined as a subset of $\{0,1\}^n$ of size $M$ such that every element has the same Hamming weight $w$. The parameter $d$ is the minimum distance of the code defined as
\bean
d & = & \min_{\substack{{\bf c}_1, {\bf c}_2 \in {\cal C} \\ {\bf c}_1 \neq {\bf c}_2}} d_H({\bf c}_1, {\bf c}_2)
\eean
where $d_H({\bf c}_1, {\bf c}_2)$ denotes the Hamming distance between the binary vectors ${\bf c}_1, {\bf c}_2$. The function $A(n,d,w)$ is the maximum possible size $M$ of a binary constant weight code of blocklength $n$, weight $w$ and minimum distance $d$. When $d=2$, there is no additional constraint on the codebook and therefore it is clear that 
\bea \label{eq:an2w}
A(n,2,w) & = & {n \choose w} .
\eea 
While there is a rich body of literature that attempt on characterizing  $A(n,d,w)$ for $d \geq 4$  \cite{Joh62,GraS80,BroSSS90,AgrVZ00, aeb,MorZKZ95, BitE95}, it still remains open in the general setting. 

Along with characterization of $A(n,d,w)$, another pertinent problem in the field of constant weight codes is the design of such codes that admit fast implementation of encoding and decoding. Considering the ease of implementation using digital hardware, it is desirable that the encoding algorithm takes in fixed-length binary vectors as input. In many systems employing a binary constant weight code, only a subset of the codebook having size as a power of $2$ is used to enable efficient implementation, and the rest of the codebook is ignored (e.g., see \cite{NorV03}). Therefore we constrain the size of the codebook to $M=2^k$ for some positive integer $k$. We refer to $k$ as the {\em combinatorial dimension} of the code. The design of low-complexity algorithms for encoding and decoding constant weight codes has been posed as a problem (Research Problem $17.3$) in the widely recognized textbook by MacWilliams and Sloane \cite{MacSloane}. In the present paper, we focus on this problem for the simplest case of $d=2$ assuming a codebook size of $M=2^k$, with an aim to achieve the largest possible $k$. 

Since $d=2$, any binary vector of weight $w$ can be included in the codebook and therefore our problem of interest aligns with the problem considered by Schalwijk \cite{Sch72} to enumerate all binary $n$-sequences of weight $w$. In \cite{Cov73}, Cover generalized Schalwijk's indexing scheme to make it applicable to an arbitrary subset of $n$-sequences. Prior to the works of Schalwijk and Cover, the indexing of constant weight $n$-sequences of weight $w$ was studied in combinatorial literature; for example, Lehmor code \cite{Leh60} produces an indexing different from that of Schalwijk's scheme. In combinatorial literature, an $n$-sequence of weight $w$ is identified as a $w$-subset (or $w$-combination) of $\{0,1,\ldots, n-1\}$ and the set of all $w$-combinations is assigned with an order, for instance the lexicographic order. The rank of a $w$-subset ${S}$ is the number of $w$-subsets that are strictly less than $S$ with respect to the lexicographic order, and the set $S$ is indexed using its rank. A procedure to compute the rank of a $w$-subset is referred to as a ranking algorithm and conversely, to recover the $w$-subset associated to a given rank as an unranking algorithm. The study of ranking/unranking algorithms and their complexity dates back to \cite{Pas87}. There are many unranking algorithms \cite{Kno74, Er85, Kok95, RusW09, GenP21, KruSKR22} proposed in literature aimed primarily at reducing the time complexity. However, all these algorithms require costly computation of binomial coefficients that have either large time complexity if done online or space complexity in case these coefficients are precomputed and stored in lookup tables. The first attempt to avoid computation of binomial coefficients is made by Sendrier in \cite{Sen05}, but the resulting code is of variable blocklength. Given this background, our paper makes the following contributions.

\ben
\item 
We present a family of binary $(n,M=2^k,d=2)$ constant weight codes ${\cal C}[\ell]$ parameterized by an integer $\ell \geq 3$. The code has blocklength $n=2^{\ell}$, weight $w = \ell$ and combinatorial dimension $k=k_{\ell}$ as defined in \eqref{eq:kl}. The code admits an encoding algorithm (Algorithm~\ref{alg:g1}) that is of linear complexity in input size $k_{\ell}$. Except for the linear time-complexity spent on parsing the input, its decoding algorithm (Algorithm~\ref{alg:g2}) has a time-complexity that is poly-logarithmic in input size $n$. Neither the encoding nor the decoding require computation of binomial coefficients.
\item The code ${\cal C}[\ell]$ is associated to a finite integer sequence $f_{\ell}$ of length $\ell$ defined in Definition~\ref{defn:fl} that satisfies a constraint referred to as anchor-decodability that is instrumental in realizing encoding and decoding algorithms of very low complexity. Among all the codes generated by anchor-decodable sequences, we prove that ${\cal C}[\ell]$ maximizes the combinatorial dimension. At the same time, we also show that ${\cal C}[\ell]$ is not a unique code that maximizes the combinatorial dimension. This is done by providing a second code construction ${\cal \hat{C}}[\ell]$ with an alternate low-complexity decoder, but with the same combinatorial dimension as that of ${\cal C}[\ell]$ when $3 \leq \ell \leq 7$.
\item 
While the code ${\cal C}[\ell]$ has a natural price to pay in its combinatorial dimension $k$, it performs fairly well against the information-theoretic upper bound $\lfloor \log_2 A(n,2,w) \rfloor$. When $\ell =3$, it in fact achieves the upper bound, and when $\ell=4$, it is one bit away from the upper bound. In general, while both $k_{\ell}$ and $\lfloor \log_2 A(2^{\ell},2,\ell) \rfloor$ grow quadratically with $\ell$, the difference $\Delta(\ell) = \lfloor \log_2 A(2^{\ell},2,\ell) \rfloor -  k_{\ell}$ is upper bounded by  $(1+\log_2 e) \ell - 1.5\log_2 \ell$, i.e., growing only linearly with $\ell$. 
\item 
Without compromising on complexity, we derive new codes permitting a larger range of parameters by modifying ${\cal C}[\ell]$ in three different ways. In the first approach, the derived code ${\cal C}_t[\ell]$ has blocklength $n=2^\ell$, weight $w=t$ and combinatorial dimension $k$ as defined in \eqref{eq:clt} for $\log_2 t < \ell-1$. In the second approach, the derived code ${\cal D}_t[\ell]$ has blocklength $n=2^\ell$, weight $w=t$ and and combinatorial dimension $k$ as defined in \eqref{eq:dlt} for $1 \leq t \leq \ell-1$. In the third approach, the derived code ${\cal B}_t[\ell]$ has blocklength $n=2^\ell-2^t +1$, weight $w=\ell$ and combinatorial dimension $k=k_{\ell} - 2t$. For certain selected values of parameters, these codes also achieve the corresponding upper bound on $k$.
\een 


\section{The Main Code Construction \label{sec:main}}

Let $|{\bf x}|$ denote the length of a vector (or a finite sequence) ${\bf x}$. We use ${\bf x}_1 \Vert {\bf x}_2$ to denote the concatenation of two vectors ${\bf x}_1, {\bf x}_2$. Entries in a vector ${\bf x}$ of length $|{\bf x}|={\sf len}$ are denoted by $x[0],x[1],\ldots, x[{\sf len}-1]$. 
We use ${\bf x}[a,m]$ to denote the sub-vector $[x[a], \ x[(a+1) \mod {\sf len} ],$ $\cdots \ x[(a+m-1) \mod {\sf len}]]$, where the $1\leq m \leq {\sf len}$ elements are accessed in a cyclic manner starting from $x[a]$. A complementary sub-vector of length $({\sf len}-m)$ can be obtained by deleting ${\bf x}[a,m]$ from ${\bf x}$ and it is denoted by $\Bar{{\bf x}}[a,m]$. We use $\text{dec}({\bf x})$ to denote the decimal equivalent of the binary vector ${\bf x}$ assuming big-endian format (least significant bit at the far end on the right). The Hamming weight of a vector ${\bf x}$ is denoted by $w_H({\bf x})$. For integers $a, b$, we use $[a]$ to denote $\{1,2,\ldots, a\}$ and $[a \ b]$ to denote $\{a, a+1, \ldots, b\}$. We use $1^m$ to denote a vector of $m$ $1$'s and $0^m$ to denote a vector of $m$ $0$'s. We use $\text{Im}(f)$ to denote image of a function $f$. 


Our main idea behind the construction is to divide the message vector $\bx$ into $\ell$ blocks of non-decreasing lengths, and then use the decimal value of each block to determine the position of the next 1-entry in the codeword of length $2^\ell$. Following this rule, the gaps among the $\ell$ 1-entries in a codeword will also allow us to recover the message uniquely. We first start with a simple warm-up construction in Section~\ref{subsec:warm-up}, which provides the intuition behind our approach, before developing the general construction and related theorems in Sections~\ref{subsec:seq1},~\ref{subsec:encoding}, and~\ref{subsec:decoding}.

\subsection{A Warm-Up Construction}
\label{subsec:warm-up}

Let us restrict that $\ell$ is a power of $2$. The encoding works as follows. First, we divide the binary message vector $\bx$ into $\ell$ blocks $\bxl,\bxlmo,\bxlmt,\ldots,\bxt,\bxo$ of lengths $\ell,\ell-\log_2 \ell ,\ell-\log_2 \ell,\ldots,\ell -\log_2 \ell,\ell -\log_2 \ell-1$, respectively without altering the order of bits, i.e., $\bx=\bxl||\bxlmo||\bxlmt||\ldots||\bxt||\bxo$. For instance, with $\ell=4$, we will have the sequence $1,2,2,4$ such that $i$-th element of the sequence is the length of ${\bf x}_i$ for $i=1,2,3,4$. With $\ell=8$, we have the sequence $4,5,5,5,5,5,5,8$. Note that the length of $\bx$ is $|\bx|=\ell+(\ell-1)(\ell-\log_2\ell)+(\ell-\log_2\ell-1)=\ell^2 - \ell\log_2\ell + (\log_2\ell-1)$. 

Next, we encode this message into a binary codeword $\bc$ of length $2^\ell$ and Hamming weight $\ell$ as follows. We set $\bc=(c[0],c[1],\ldots,c[2^\ell-1])$ to the all-zero codeword and index its bits from 0 to $2^\ell-1$.
Let $\pos_{\ell}\triangleq \dec(\bxl)$ be the decimal value of the block $\bxl$. Leave the first $\pos_{\ell}$ bits unchanged as 0's, but set the $(\pos_{\ell}+1)$-th bit of $\bc$ to one, i.e. $c[\pos_{\ell}] \triangleq 1$. Now, we move to $\bxlmo$ and again let $\pos_{\ell-1} \triangleq \dec(\bxlmo)$. We skip $\pos_{\ell-1}$ 0's after the first 1, and set the next bit to 1, i.e. 
$c[(\pos_\ell+\pos_{\ell-1}+1)\mod 2^\ell]\triangleq 1$. Note that here we move from the left to the right cyclically along the codeword indices, wrapping around at the end. We continue the process until the last block $\bxo$ is read and the last 1 is add to $\bc$.

\begin{figure}[htb!]
    \centering
    \includegraphics[scale=0.78]{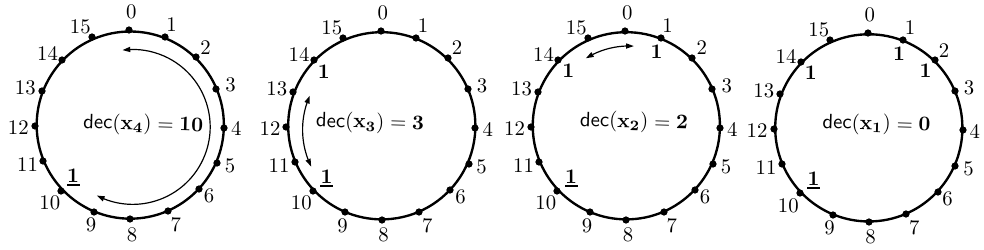}
    \caption{Illustration of the encoding process when $\ell=4$ and the message vector $\bx=(1,0,1,0,1,1,1,0,0)$ is encoded into the codeword $\bc$ of length $16=2^4$ (represented by the circle) with $c[1]=c[2]=\bc[10]=c[14]=1$. For decoding, one first determine the \textit{anchor} (the underlined 1), which is the 1 that has the largest number of consecutive zeros on its left (cyclically), or equivalently, has the largest gap to the nearest 1 on its left. Once the anchor is found, each message block can be recovered by counting the number of 0's between the current 1 to the next.}
    \label{fig:warm-up}
\end{figure}

For the example illustrated in  Fig.~\ref{fig:warm-up}, when $\ell=4$, the message vector $\bx = (1,0,1,0,1,1,1,0,0)$ is divided into $\bx_4=(1,0,1,0)$, $\bx_3=(1,1)$, $\bx_2=(1,0)$, and $\bx_1=(0)$, which are of lengths $4, 2, 2, 1$ as described earlier. Since $\dec(\bx_4)=10$, we set $c[10]=1$, noting that the bits of $\bc$ are indexed from 0 to 15. Next, since $\dec(\bx_3)=3$, we set $c[14]=c[(10+3+1)]=1$. Similarly, as $\dec(\bx_2)=2$ and $\dec(\bx_1)=0$, we set $c[1] = c[14+2+1] = 1$ and $c[2] = c[1+0+1] = 1$. As the result, $\bc = (0,1,1,0,0,0,0,0,0,0,\underline{1},0,0,0,1,0)$. To decode, given such a codeword $\bc$, we need to reconstruct $\bx$. Clearly, if the position of the ``first'' 1 (called the \textit{anchor}), which corresponds to the block $\bxl$ is known, then $\bxl$ can be recovered right away. Moreover, the gap (that is, the number of 0's) between this 1 and the next 1 on its right (cyclically, wrapping around if necessary) will be the decimal value of the block $\bxlmo$. 
For example, if we know the 1 at index 10 of $\bc$ (the underlined one) is the anchor, then we can derive immediately that $\bx_4=(1,0,1,0)$. Moreover, we can simply count the number of 0's between this 1 and the next, which is 3, and recover $\bx_3 = (1,1)$. 
All the $\ell$ blocks of $\bx$ can be recovered in this way. Thus, the key step is to determine the anchor. 

We claim that thanks to the way we split $\bx$, the 1 with the \textit{largest} number of 0's on its left (wrapping around if necessary) in $\bc$ is the anchor, created by $\bxl$. Note that for the 1's created by $\bxo,\ldots,\bxlmo$, the numbers of 0's on their left are at most $\max_{\bxlmo}\dec(\bxlmo) = 2^{\ell-\log_2\ell}-1 = \frac{2^{\ell}}{\ell} - 1$. 
On the other hand, for every $\ell\geq 3$, the number of 0's on the left of the anchor is at least
\begin{align}
\label{eq:condition1} 
2^\ell - \ell - \bigg(\sum_{i=1}^{\ell-2} (2^{\ell-\log_2\ell}-1) + (2^{\ell-\log_2\ell-1}-1) \bigg) = \frac{(\tfrac{3}{2}) \cdot 2^{\ell}}{\ell} - 1 \ \geq \ \frac{2^{\ell}}{\ell} > \frac{2^{\ell}}{\ell}-1,
\end{align}
which proves our claim.

Finally, note that this warm-up construction assumes $\ell$ as a power of $2$. This can be generalized for any $\ell \geq 3$.

\subsection{A Finite Integer Sequence\label{subsec:seq1}}

In this subsection we generalize the sequence used in the warm-up construction for every $\ell \geq 3$.

\bdefn \label{defn:fl} Let $\ell \geq 3$. Then $f_{\ell}(i), i = 1,2,\ldots, \ell$ is a finite integer sequence of length $\ell$ defined as follows. If $\ell$ is not a power of $2$, then 
\bea \label{eq:fl1}
f_{\ell}(i) = \left\{ \begin{array}{ll}
      \ell - \lceil \log_2 \ell \rceil, &  \ \ i = 1, 2, \ldots, \ell-\mu \\ 
        \ell - \lfloor \log_2 \ell \rfloor,   & \ \ i = \ell-\mu+1, \ell-\mu+2, \ldots, \ell-1   \\
        \ell,   & \ \ i = \ell 
\end{array} \right. 
\eea 
where $\mu = 2^{\lceil \log_2 \ell \rceil} - \ell$. If $\ell$ is a power of $2$, then 
\bea \label{eq:fl2}
f_{\ell}(i) = \left\{ \begin{array}{ll}
       \ell - \log_2 \ell - 1, &  \ \ i = 1 \\
        \ell - \log_2 \ell ,   & \ \ i = 2, 3, \ldots, \ell-1 \\
        \ell,   & \ \ i = \ell 
\end{array} \right. . 
\eea 
\edefn 
Next we define
\bea 
\label{eq:kl} k_{\ell} & \triangleq & \sum_{i=1}^\ell f_{\ell}(i)  \\ 
\label{eq:klval} & = &  \left\{ \begin{array}{ll}
        \ell^2 - (\mu \lfloor \log_2\ell \rfloor +(\ell-\mu)\lceil \log_2\ell \rceil) +  \lfloor \log_2\ell \rfloor ,   & \ \ \ell \text{ is not a power of 2} \\
        \ell^2 - \ell \log_2 \ell + (\log_2 \ell - 1),   & \ \ \ell \text{ is a power of 2}
\end{array} \right. . 
\eea
The lower bound on $k_{\ell}$ obtained in the following proposition gives a lucid estimate on how it grows with $\ell$. 
\bprop \label{prop:klbound} Let $\ell \geq 3$ be an integer. Suppose $\ell = 2^a+b$ such that $2^a \leq \ell$ is the maximum power of $2$ and $b\geq 0$. Then
\bea \label{eq:kllb}
k_{\ell} & \geq & \left\{ \begin{array}{ll}
     \ell^2 - \ell \log_2\ell + \log_2 \ell - 1, & b = 0 \\
     \ell^2 - \ell \log_2\ell + \log_2 \ell - b\bigl(2-\tfrac{1}{\ln 2}\bigr) - \bigl(\tfrac{b}{\ell}\bigr) \tfrac{1}{\ln 2}, &  b \neq 0 
\end{array} \right . 
\eea 
As a corollory, $k_{\ell} \ \geq \ \ell^2 - \ell \log_2\ell + \log_2 \ell - \ell(1-\tfrac{1}{2\ln 2}) - \tfrac{1}{2\ln 2} $ for every $\ell \geq 3$.
\eprop 
\bpf The bound in \eqref{eq:kllb} is trivially true with equality when $b=0$ and hence it is tight. When $\ell$ is not a power of $2$, i.e., $b \neq 0$, we substitute value of $\mu$ in \eqref{eq:klval} to obtain
\bea
\nonumber k_\ell & = & \ell^2 - (\ell -1) \lfloor\log_2\ell \rfloor - 2 (\ell - 2^{\lfloor \log_2\ell \rfloor}) \\
\label{eq:kllb1} & \geq & \ell^2 - (\ell-1)\Bigl(\log_2 \ell - \frac{b}{\ell\ln 2}\Bigr) - 2b  \\
\nonumber & = & \ell^2 - \ell \log_2 \ell + \log_2 \ell -  b\bigl(2-\tfrac{1}{\ln 2}\bigr) - \bigl(\tfrac{b}{\ell}\bigr) \tfrac{1}{\ln 2} 
\eea 
In \eqref{eq:kllb1}, we use an upper bound for $\lfloor \log_2 \ell \rfloor$ in terms of $\log_2 \ell $ obtained by invoking the inequality $\ln (1+x) \geq \frac{x}{1+x}$. Observe that $b < \tfrac{\ell}{2}$. We substitute it in \eqref{eq:kllb} and observe that $\ell(1-\tfrac{1}{2\ln 2}) - \tfrac{1}{2\ln 2} \geq 1$ for every $\ell \geq 3$. This proves the corollary. 
\epf

\subsection{Encoding Information in Gaps}
\label{subsec:encoding}

In this section, we present an encoding algorithm (see Algorithm~\ref{alg:g1}) that encodes information in gaps between successive $1$'s of a binary vector of length $n=2^{\ell}$, using the sequence $s_{\ell}=s_{\ell}(1),s_{\ell}(2),\ldots,s_{\ell}(\ell)$ where $s_{\ell}(\ell)$ is fixed to be $\ell$. More specifically, the message vector $\bx$ will be divided into $\ell$ blocks $\bxl,\ldots,\bxt, \bxo$, which are of lengths $s_{\ell}(\ell),\ldots,s_{\ell}(2), s_{\ell}(1)$, and gaps between successive $1$'s of the codewords depend on the decimal value of each of these blocks. The function {\sf gap} defined below formalizes the notion of gap as the first step.
\bdefn \label{defn:gap} Let $a, b \in \mathbb{Z}_n$. Then the gap from $a$ to $b$ is a natural number taking values in $[0 \ (n-1)]$ given by 
\bean
{\sf gap}(a,b) = (b-a-1) \mod n .
\eean 
\edefn

The encoding algorithm given in Algorithm~\ref{alg:g1} is invoked taking the sequence $s_{\ell}$ as an auxiliary input. The input ${\bf x}$ is the message vector that gets encoded, and its length must be 
\bea \label{eq:ksl}
k(s_{\ell}) & \triangleq & \sum_{i}s_{\ell}(i) .
\eea
The encoded vector is the output ${\bf c}$ of length $n$. The input vector ${\bf x}$ is partitioned as ${\bf x}_{\ell}\Vert {\bf x}_{\ell-1} \Vert \cdots \Vert {\bf x}_{1}$ such that $|{\bf x}_{i}|=s_{\ell}(i)$ for $i \in [\ell]$. The vector ${\bf c}$ is initialized as all-zero vector and $\ell$ locations of ${\bf c}$ are set to $1$ subsequently. 
The input bits are read in blocks ${\bf x}_{\ell-1}, {\bf x}_{\ell-2},\ldots {\bf x}_{1}$ and every time a block ${\bf x}_i, \ell \geq i \geq 1$ is read, a bit in ${\bf c}$ is set to $1$ in a such manner that the gap from the previously set $1$ is equal to $\text{dec}({\bf x}_j)$.
The gap is always computed modulo $n$ so that the position pointer ${\sf pos}$ can wrap around cyclically. The algorithm has a linear time-complexity in input size $k(s_{\ell})$, and it defines the encoding map $\phi:\{0,1\}^{k(s_{\ell})} \xrightarrow{} \{0,1\}^{n}$. 
\begingroup


    

    



    


\begingroup
\begin{algorithm}
	\caption{\textsc{Encode $\phi(\cdot)$} \newline {\bf Input}: ${\bf x} \in \{0,1\}^{\sum_i s_{\ell}(i)}$, $s_{\ell}$ \newline {\bf Output}: ${\bf c} \in \{0,1\}^n$ \label{alg:g1}}
	\DontPrintSemicolon
	Partition ${\bf x}$ as ${\bf x}_{\ell}\Vert {\bf x}_{\ell-1} \Vert \cdots \Vert {\bf x}_{1}$ such that $|{\bf x}_{i}|=s_{\ell}(i)$ for $i \in [\ell]$.

    Initialize array ${\bf c}=0^n$ 
    
    ${\sf pos} \leftarrow -1$

	\For{$j = \ell,\ldots,1$}
	{ 

    ${\sf pos} \xleftarrow{} {\sf pos} + 1 + \text{dec}({\bf x}_j) \mod n$

     $c[{\sf pos}] \xleftarrow{} 1$

	}
\end{algorithm}
\endgroup
Choosing the auxiliary input $s_{\ell}$ as $f_{\ell}$ defined in Sec.~\ref{subsec:seq1} and fixing $\ell=4$ recovers the warm-up construction presented in Sec.~\ref{subsec:warm-up}. Apart from the fact that $s_{\ell}(\ell)=\ell$ always, there is room to vary $s_{\ell}(i),i=1,2,\ldots, \ell-1$. Thus Algorithm~\ref{alg:g1} provides a generic method to encode information as gaps in a vector of length $n=2^{\ell}$. What it requires is to identify a ``good'' sequence so as to produce a code that is easily decodable and at the same time has high combinatorial dimension. 

\subsection{A Decodability Criterion and a Decoding Algorithm }
\label{subsec:decoding}

In this subsection, we first establish a criterion for unique decodability of a vector ${\bf c}$ obtained as the output of the encoding algorithm $\phi$. The criterion solely depends on the auxiliary input $s_{\ell}$ and is stated in Definition~\ref{defn:udec}.


\bdefn \label{defn:shift} Let ${\bf g} = (g[0], g[1],\ldots, g[\ell-1])$ be a vector of length $\ell$. Then the circular shift of ${\bf g}$ by  $\ell_0 \in \mathbb{Z}_{\ell}$ is defined as 
\bea \label{eq:cshift}
{\sf cshift}({\bf g},\ell_0) & = & (g[\ell_0],g[\ell_0+1],\ldots, g[\ell-1],g[0],\ldots, g[\ell_0-1]) .
\eea 
For any $\ell_0 \in \mathbb{Z}$, the definition still holds true by replacing $\ell_0$ by $\ell_0 \mod \ell$  in \eqref{eq:cshift}.
\edefn

\bdefn \label{defn:udec} Let $\ell \geq 3$ be an integer. A non-decreasing sequence $s_{\ell}$ of length $\ell$ is said to be anchor-decodable if $s_{\ell}(\ell)=\ell$ and the following two conditions hold:
\ben  
    \item \bea \label{eq:udec1}
2^{\ell} - \sum_{i=1}^{\ell-1} 2^{s_{\ell}(i)} & \geq & 2^{s_{\ell}(\ell-1)}. 
\eea 
    \item The vector 
    $\boldsymbol{\gamma} = (2^{\ell} - 1 - \sum_{i=1}^{\ell-1} 2^{s_{\ell}(i)}, 2^{s_{\ell}(\ell-1)}-1,2^{s_{\ell}(\ell-2)}-1,\ldots,2^{s_{\ell}(1)}-1)$
    is distinguishable from any of its cyclic shifts, i.e., ${\sf cshift}(\boldsymbol{\gamma},\ell_0) \neq \boldsymbol{\gamma}$ for every integer $0 < \ell_0 < \ell$.
\een 
\edefn 
In what follows in this subsection, we will describe why the conditions in Defn.~\ref{defn:udec} are important and how they naturally lead to a fast decoding algorithm as presented in Algorithm~\ref{alg:g2}. As the first step, we show that the Hamming weight of $\phi({\bf x})$ is always $\ell$ for every input ${\bf x}$ to the encoder in Alg.~\ref{alg:g1} if the sequence $s_{\ell}$ is anchor-decodable.

\blem \label{lem:wt} Let $\ell\geq 3$ and $n = 2^\ell$. If $s_{\ell}=\big(s_{\ell}(1),s_{\ell}(2),\ldots,s_{\ell}(\ell)\big)$ is an anchor-decodable sequence, then $w_H(\phi({\bf x})) = \ell$ for every ${\bf x} \in \{0,1\}^{k(s_{\ell})}$, where $\phi(\cdot)$ is determined by Alg.~\ref{alg:g1}.
\elem 
\bpf Let ${\bf c} = \phi({\bf x})$. After completing the first iteration of the loop in \textsl{Line} $4$ of Alg.~\ref{alg:g1}, the position pointer ${\sf pos}$ takes a value $p_0=\text{dec}({\bf x}_{\ell})$ lying between $0$ and $2^\ell-1$, and ${\bf c}$ has Hamming weight $1$ with $c[p_0]=1$. The loop has $(\ell-1)$ remaining iterations indexed by $j=\ell-1,\ldots, 1$. In each of these $(\ell-1)$ iterations, ${\sf pos}$ is incremented modulo $n$ at least by $1$ and at most by $2^{|{\bf x}_j|}, j \in [\ell-1]$. Therefore, the maximum cumulative increment $p$ in ${\sf pos}$ from $p_0$ by the end of these  $(\ell-1)$ iterations is given by:
\bean 
\nonumber p & = & \sum_{j=1}^{\ell-1} 2^{|{\bf x}_j|}  \ = \ \sum_{j=1}^{\ell-1} 2^{s_{\ell}(j)} 
\eean
If $s_{\ell}$ is anchor-decodable, then from \eqref{eq:udec1} we obtain that
\bea \label{eq:p}
p & \leq & 2^{\ell} - 2^{s_{\ell}(\ell-1)} \ < \ 2^{\ell}.
\eea 
Since $p < n$, a distinct bit of ${\bf c}$ is flipped from $0$ to $1$ in every iteration and therefore $w_H({\bf c})=\ell$.
\epf
Let us view the input ${\bf x}$ as concatenation of $\ell$ binary strings as ${\bf x} = {\bf x}_{\ell} \Vert {\bf x}_{\ell-1} \Vert \cdots \Vert {\bf x}_{1}$ where $|{\bf x}_i|=s_{\ell}(i)$. Suppose that ${\bf c} = \phi({\bf x})$ is the output of Alg.~\ref{alg:g1}. By Lemma~\ref{lem:wt}, ${\bf c}$ has $\ell$ $1$'s. Let $j[m], m=0,1,\ldots, \ell-1$ denote the locations of $1$'s in ${\bf c}$ counting from left to right and let 
\bea \label{eq:gapvector}
\gaps[m] = \gap(j[(m-1) \mod \ell], j[m]), \ \ m=0,1,\ldots \ell-1 .
\eea  
denote the array of the number of zeros between two successive $1$'s cyclically wrapping around ${\bf c}$ if required. The principle of the decoding algorithm in Algorithm~\ref{alg:g2} is to uniquely identify the anchor bit of ${\bf c}$ assuming that the sequence $s_{\ell}$ is anchor-decodable. Recall (Sec.~\ref{subsec:warm-up}) that the anchor bit in a codeword ${\bf c}$ is the first bit flipped to $1$ while running the encoding algorithm to generate ${\bf c}$. To be precise, 
\bea \label{eq:init}
j[{\sf anchor\_index}] & = & j \text{ such that } c[j] \text{ is the first bit set to $1$ while encoding } {\bf c} 
\eea 
and we call $j[{\sf anchor\_index}]$ as the anchor and ${\bf c}[j[{\sf anchor\_index}]]$ as the anchor bit $1$. The procedure \textsc{FindAnchor} (Algorithm~\ref{alg:g3}) invoked at \textsl{Line} $3$ of Alg.~\ref{alg:g2} returns ${\sf anchor\_index}$ and its correctness will be analyzed shortly. If the index ${\sf anchor\_index}$ is uniquely identified by an input vector ${\bf c}$, then it is straightforward to observe that ${\bf x}_{\ell}, {\bf x}_{\ell-1}, \ldots, {\bf x}_{1}$ are uniquely determined. The procedure to recover ${\bf x}$ given the knowledge of $j[{\sf anchor\_index}]$ is laid down in \textsl{Lines} $4-8$ of Algorithm~\ref{alg:g2}. 

\vspace{5pt}
\begingroup
\begin{algorithm}[H]
	\caption{\textsc{Decode}  \newline {\bf Input}: ${\bf c} \in \text{Im}(\phi), s_{\ell} $ \newline {\bf Output}: ${\bf x} \in \{0,1\}^{k(s_{\ell})}$ \label{alg:g2}}
	\DontPrintSemicolon
	Find $0 \leq j[0] < j[1] < \cdots < j[\ell -1 ] < n$ such that $c[j[i]] = 1$ for every $i=0,1,\ldots, \ell-1$. 
    
    $\gaps[m] = {\sf gap}(j[(m-1) \hspace{-1.5mm}\mod \ell], j[m])$ for $m=0,1,\ldots \ell-1$

    ${\sf anchor\_index} = \textsc{FindAnchor}({\bf g}, s_{\ell})$
    
    Initialize binary vector ${\bf x}$ such that $|{\bf x}|=\ell$ and $\text{dec}({\bf x})=j[{\sf anchor\_index}]$
    
    \For{$i = 1,2,\ldots, \ell-1$}    
    {
		$g \leftarrow \gaps[({\sf anchor\_index}+i) \hspace{-1.5mm}\mod \ell]$
		
		Represent $g$ as binary string ${\bf x}_{i}$ of length $s_{\ell}(\ell-i)$

		${\bf x} \leftarrow {\bf x}\Vert {\bf x}_{i}$
    }
\end{algorithm}
\endgroup
\vspace{5pt}

\begingroup
\begin{algorithm}[H]
	\caption{\textsc{FindAnchor}  \newline {\bf Input}: ${\bf g} \in \mathbb{Z}_n^{\ell}, s_{\ell}$ \newline {\bf Output}: ${\sf anchor\_index} \in [0 \ \ell-1]$ \label{alg:g3}}
	\DontPrintSemicolon
    ${\sf gaps\_allone} \leftarrow (2^{\ell}-1-\sum_i 2^{s_{\ell}(i)}) \Vert (2^{s_{\ell}(i)}-1,i=\ell-1,\ell-2,\ldots, 1)$
    
    \If{ $\exists n_0 \in \mathbb{Z}_{\ell}$ \emph{such that} ${\sf gaps\_allone} = {\sf cshift}({\bf g},n_0)$ }
    {
        ${\sf anchor\_index} \leftarrow n_0$
        
    } \Else
    {
        ${\sf anchor\_index} = \arg\max_m \{{\bf g}[m] \mid m = 0,1,\ldots, \ell-1\}$
    }
\end{algorithm}
\endgroup
\vspace{5pt}

Let us proceed to check the correctness of Algorithm~\ref{alg:g3} \textsc{FindAnchor}. It is straightforward to see that:
\bea \label{eq:gapsum} 
n & = & \ell + \sum_{m=0}^{\ell-1} \gaps[m] \ = \ \ell + \gaps[{\sf anchor\_index}] + \sum_{i = 1}^{\ell-1} \gaps[({\sf anchor\_index}+i ) \hspace{-2.5mm}\mod \ell] .
\eea 
Therefore we have
\bea 
\nonumber \gaps[{\sf anchor\_index}]  & = & (n - \ell) - \sum_{i = 1}^{\ell-1} \gaps[({\sf anchor\_index}+i ) \hspace{-2.5mm}\mod \ell] \\
 \label{eq:initlb11} & \geq & (n - \ell) - \sum_{i = 1}^{\ell-1} (2^{|{\bf x}_{\ell-i}|} - 1) \\ 
 \nonumber & = & (2^{\ell} - \ell) - \sum_{i = 1}^{\ell-1} (2^{s_{\ell}(\ell -i)} - 1) \ = \ 2^{\ell} - 1 - \sum_{i = 1}^{\ell-1} 2^{s_{\ell}(\ell -i)} .
\eea 
The inequality in \eqref{eq:initlb11} follows from the way ${\bf x}_{\ell-i}$ is encoded by Algorithm~\ref{alg:g1}. It is straightforward to check that equality holds in \eqref{eq:initlb11} if and only if the message vector is of the type
\bea \label{eq:allone1}
{\bf x}_{\ell - i} & = & 1^{s_{\ell}(\ell-i)}, \text{~~for all}~ i=1,2,\ldots, \ell-1. 
\eea 
When the message vector satisfies \eqref{eq:allone1}, every gap except $\gaps[{\sf anchor\_index}]$ becomes maximal in length, and therefore we refer to this special case as the {\em maximal-gap} case. The \textsl{Lines} $2-3$ in Alg.~\ref{alg:g3} check for the maximal-gap case by comparing every circular shift of the vector ${\bf g}$ with a fixed vector ${\sf gaps\_allone}$. The vector ${\sf gaps\_allone}$ corresponds to a message vector of the type
\bea \label{eq:gapallone}
{\bf x}_{i} & = & 1^{s_{\ell}(i)},\text{~for all}~ i=1,2,\ldots, \ell-1, \text{~~and}~~  \text{dec}({\bf x}_{\ell}) \ \leq \ 2^{\ell} - 1 - \sum_{i = 1}^{\ell-1} 2^{s_{\ell}(\ell -i)}  
\eea 
for which ${\sf anchor\_index} = 0$. If ${\sf cshift}({\bf g},n_0)$ becomes equal ${\sf gaps\_allone}$ for some $0 \leq n_0 \leq (\ell-1)$, then by second condition in Defn.~\ref{defn:udec}, $n_0$ is unique and is equal to ${\sf anchor\_index}$. 

If  \eqref{eq:allone1} is false, then clearly \eqref{eq:initlb11} satisfies with strict inequality, and in that case 
\bea
\gaps[{\sf anchor\_index}] & > & 2^{\ell} - 1 - \sum_{i = 1}^{\ell-1} 2^{s_{\ell}(\ell -i)} \\
& \geq & (2^{s_{\ell}(\ell-1)} - 1) \ \geq \ \max_{i=1,\ldots,\ell-1} (2^{s_{\ell}(i)} - 1) .
\eea
by the first condition of Defn.~\ref{defn:udec} and the fact that $s_{\ell}$ is non-decreasing. Thus \textsl{Line} $5$ of Algorithm~\ref{alg:g3} correctly identifies the ${\sf anchor\_index}$ and therefore the it is correct if the sequence $s_{\ell}$ is anchor-decodable. Thus Algorithm~\ref{alg:g2} provides an explicit decoder that maps ${\bf c}$ uniquely to ${\bf x}$ leading to the following theorem.

\bthm \label{thm:code} Let $\ell\geq 3$ and $n=2^\ell$. For every anchor-decodable sequence $s_{\ell}$ as defined in Definition~\ref{defn:udec}, the map $\phi$ defined by Algorithm~\ref{alg:g1} with $s_{\ell}$ as auxiliary input is one-to-one. Furthermore, for every ${\bf x} \in \{0,1\}^{k(s_{\ell})}$ with $k(s_{\ell})=\sum_i s_{\ell}(i)$, Algorithm~\ref{alg:g2} outputs ${\bf x}$ when $\phi({\bf x}) \in \{0,1\}^{2^{\ell}}$ is passed as its input.  
\ethm 

    

    
    
		


    
        

\subsection{Constant Weight Codes\label{subsec:code}}

By Theorem~\ref{thm:code} and Lemma~\ref{lem:wt}, every anchor-decodable sequence $s_{\ell}$ has an associated binary constant weight code $\phi(\{0,1\}^{k(s_{\ell})})$. We define
\bea \label{eq:cs;}
{\cal C}[s_{\ell}] & \triangleq & \phi(\{0,1\}^{k(s_{\ell})})
\eea 
and call $s_{\ell}$ as the {\em characteristic sequence} of ${\cal C}[s_{\ell}]$. The codewords of ${\cal C}[s_{\ell}]$ can be obtained as $2^{k(s_{\ell})}$ distinct permutations of $1^{\ell}\Vert 0^{n-\ell}$. This is a subcode of Type I permutation modulation of size $2^{\ell}\choose \ell$, with initial vector $1^{\ell}\Vert 0^{n-\ell}$ introduced in \cite{Sle65}. Therefore the encoder $\phi$ gives an elegant method to map binary vectors of length $k(s_{\ell})$ to a subset of the permutation code, which is otherwise usually carried out by picking vectors in lexicographic order~\cite{NorV03}. 

In the following, we verify that $f_{\ell}$ defined in Defn.~\ref{defn:fl} is an anchor-decodable sequence. Clearly $f_{\ell}$ is non-decreasing and $f_\ell(\ell)=\ell$. When $\ell$ is not a power of $2$,
\bea 
\nonumber \sum_{i=1}^{\ell-1} 2^{f_{\ell}(i)} & = & (\mu -1) 2^{\ell - \lfloor \log_2 \ell \rfloor} + (\ell-\mu) 2^{\ell - \lceil \log_2 \ell \rceil} \\
\nonumber & = & 2^{\ell} \left\{ \mu 2^{- \lfloor \log_2 \ell \rfloor} + (\ell-\mu) 2^{- \lceil \log_2 \ell \rceil}\right\} - 2^{\ell - \lfloor \log_2 \ell \rfloor} \\
\label{eq:p11} & = & 2^{\ell} - 2^{\ell - \lfloor \log_2 \ell \rfloor} \\
\label{eq:p12} & \leq &   2^{\ell} - 2^{f_{\ell}(\ell-1)} ,
\eea  
and \eqref{eq:p12} holds with equality if and only if $\mu > 1$. In the above, \eqref{eq:p11} follows by substituting the value of $\mu$ and calculating that:
\bean
\mu 2^{- \lfloor \log_2 \ell \rfloor} + (\ell-\mu) 2^{- \lceil \log_2 \ell \rceil} & = &  2\mu \cdot 2^{- \lceil \log_2 \ell \rceil} + (\ell - \mu) \cdot 2^{- \lceil \log_2 \ell \rceil} \\
& = & (\mu + \ell) \cdot 2^{-\lceil \log_2 \ell \rceil} \\
& = & (2^{\lceil \log_2 \ell \rceil} - \ell + \ell)  \cdot 2^{-\lceil \log_2 \ell \rceil} \ = \ 1. 
\eean 
 On the other hand, when $\ell$ is a power of $2$,
\bea 
\nonumber \sum_{i=1}^{\ell-1} 2^{f_{\ell}(i)} & = & \ell \cdot 2^{\ell - \log_2 \ell } - 2^{\ell - \log_2 \ell } - 2^{\ell - \log_2 \ell -1} \\
\nonumber & = & 2^{\ell} - (\tfrac{3}{2}) \cdot 2^{\ell - \log_2 \ell } \\
\label{eq:p22}  & < & 2^{\ell} - 2^{\ell - \log_2 \ell } \ = \ 2^{\ell} - 2^{f_{\ell}(\ell-1)} .
\eea  
By \eqref{eq:p12} and \eqref{eq:p22}, the first condition of anchor-decodability is satisfied. In order to check for the second condition in Defn.~\ref{defn:udec}, let us first compute $\boldsymbol{\gamma}$ as:
\bean
\boldsymbol{\gamma} & = & \left\{ \begin{array}{cc}
     \Bigl(2^{\ell} \bigl(1-\frac{3}{2\ell}\bigr)-1, \frac{2^{\ell}}{\ell}-1,\ldots, \frac{2^{\ell}}{\ell}-1, \frac{2^{\ell}}{2\ell}-1 \Bigr), &  \ell \text{ is a power of 2} \\
     \Bigl(\underbrace{\frac{2^{\ell}}{2^{\lfloor \log_2 \ell \rfloor}} -1,\ldots, \frac{2^{\ell}}{2^{\lfloor \log_2 \ell \rfloor}} -1}_{\mu \text{ terms}}, \underbrace{\frac{2^{\ell}}{2^{\lceil \log_2 \ell \rceil}} -1, \ldots, \frac{2^{\ell}}{2^{\lceil \log_2 \ell \rceil}} -1}_{\ell-\mu \text{ terms}} \Bigr), &  \text{otherwise} \\ 
\end{array} \right. .
\eean 
Clearly, $\boldsymbol{\gamma}$ is distinguishable from any of its $(\ell-1)$ non-trivial cyclic shifts as $1 \leq \mu \leq \ell-2$, establishing that $f_{\ell}$ is anchor-decodable. As will be shown in the next subsection, $f_{\ell}$ is in fact an optimal anchor-decodable sequence producing the largest possible code ${\cal C}[\ell]$ as defined  below.
\bdefn \label{defn:clcode} Let $\ell \geq 3$. We define the code ${\cal C}[\ell] = {\cal C}[f_{\ell}]$ where the characteristic sequence $s_{\ell}$ is chosen as $f_{\ell}$. The code has blocklength $n=2^\ell$, weight $w=\ell$, and combinatorial dimension $k=k(f_{\ell})=k_{\ell}$ where $k_{\ell}$ is given in \eqref{eq:kl}.
\edefn 


\subsection{On the Optimality of ${\cal C}[\ell]$\label{subsec:opt}}

In this section, our interest is to identify an anchor-decodable sequence $s_{\ell}$ that attains the maximum combinatorial dimension for its associated code ${\cal C}[s_{\ell}]$. In the following theorem, we establish that $f_{\ell}$ maximises $k(s_{\ell})$.

\bthm \label{thm:cloptimal} Let $\ell \geq 3$. Among all anchor-decodable sequences $\{s_{\ell} \}$ as defined in Definition~\ref{defn:udec}, the sequence $f_{\ell}$ as defined in Definition~\ref{defn:fl} maximizes $k(s_{\ell})=\sum_{i}s_{\ell}(i)$.
\ethm 
\bpf We first give an overview of the proof technique. Our approach is to transform the maximization problem into an equivalent problem that is related to minimization of average length of a source code for a discrete source with alphabet-size $\ell$. It is well-known that Huffman algorithm yields an optimal source code having the minimum average length. After establishing necessary  equivalences, the optimal codeword lengths of Huffman code are  made use of to construct a sequence that maximizes $k(s_{\ell}) = \sum_{i}s_{\ell}(i)$. It turns out that the resultant sequence is indeed $f_{\ell}$.

In the first step, we consider a discrete source with an alphabet ${\cal A}=\{a_1, a_2,\ldots, a_{\ell}\}$ and a uniform probability mass function, i.e., $\Pr(a_i) = (1/\ell)$ for every $i$. By slight abuse of notation, we use ${\cal A}$ to denote the source as well. A binary source code is a mapping $s:{\cal A} \xrightarrow{} \{0,1\}^*$ and we say $a_i$ has a codeword length  $L(a_i) \triangleq |s(a_i)|$. The average length of the source code is defined as
\bean
\Bar{L}({\cal A}) & = & \sum_{i \in [\ell]} \Pr(a_i) L(a_i) \ = \ \sum_{i \in [\ell]} (1/\ell) L(a_i).
\eean 
A source code that minimizes $\Bar{L}({\cal A})$ over all possible source codes is called an optimal code and it is well-known that Huffman encoding algorithm produces an optimal source code \cite{Gallager_2008}. The Huffman algorithm constructs a rooted binary tree of $\ell$ leaf nodes in which each symbol $a_i$ uniquely corresponds to a leaf node. We call it a Huffman code tree. Let $T_H = (V_H, E_H)$ be the Huffman code tree associated to the source ${\cal A}$ with root node $v_{r}$ and $\ell$ leaf nodes $v(a_i), i=1,2,\ldots , \ell$. The binary codeword associated to $a_i$ can be identified from the leaf node as follows. Among two possible children of a node $v$ in the binary tree, the edge from $v$ to the left one is marked as $0$ and to the right one as $1$. Let $P(v)$ denote the unique path from $v_{r}$ to an arbitrary node $v$. Then the unique path from $v_{r}$ to the leaf node $v(a_i) \in V_H$ identifies a binary string. This forms the codeword $s(a_i)$. The depth of a node $v$ in a binary tree is the length of the unique path $P(v)$ and is denoted by $L_T(v)$. Therefore, $L_T(v(a_i))=L(a_i)$. A source code that can be represented as a rooted binary tree with leaves representing codewords as described above is called a prefix-free code. Hence Huffman code is an optimal code that is prefix-free as well. The following two lemmas are relevant for our proof.

\blem \cite{Gallager_2008} \label{lem:Huff} Consider a discrete source with with alphabet ${\cal A} = \{a_i, i=1,2,\ldots, \ell\}$. Let $T_H$ denote the Huffman code tree of the source. Then
\ben 
\item $T_H$ is full.
\item If the source has uniform distribution, then for every leaf node $v(a_i)$, $L_{T_H}(v(a_i))$ is either $\lceil \log_2\ell \rceil$ or $\lfloor \log_2\ell \rfloor$.
\item If $\ell$ is a power of $2$ and a prefix-free source code has average length $\log_2\ell$, then $L_{T_p}(v(a_i)) = \log_2 \ell$ for every $i$ where $T_p$ is the code tree of the code.
\een 
\elem 
\bpf All these are discussed in Gallager's textbook \cite{Gallager_2008}. The first assertion is presented as Lemma~2.5.2 in \cite{Gallager_2008}. The second follows from Exercise~2.14(a) and the third from Exerice~2.9(a) in \cite{Gallager_2008}.
\epf
\blem [Kraft Inequality \cite{Gallager_2008}] \label{lem:Kraft} Consider a prefix-free source code for a discrete source with alphabet ${\cal A} = \{a_i, i=1,2,\ldots, \ell\}$. Let $L(a_1), L(a_2),\ldots, L(a_{\ell})$ denote the lengths of codewords. Then
\bea \label{eq:Kraft}
\sum_i 2^{-L(a_i)} & \leq & 1 .
\eea  
Conversely, if $L(a_1), L(a_2),\ldots, L(a_{\ell})$ are positive integers satisfying \eqref{eq:Kraft}, then there exists a prefix-free source code with these as codeword lengths. 
\elem 
In the second step, we extend the Huffman code tree $T_H$ to a form a perfect binary tree $T=(V,E)$ of depth $\ell$. For any node $v \in V$, let $N(v)$ be the set of leaf nodes of $T$ for which the unique path from $v_r$ to the leaf node includes $v$. We define $N(v)$ as the {\em canopy} of $v$. By the first property of Lemma~\ref{lem:Huff}, the canopies $N(v(a_i))$ in $T$ are pairwise disjoint and furthermore, their union forms the set of all leaf nodes of $T$. A collection of nodes $U \subset V$ is said to be {\em prefix-free} if for any $u_1, u_2 \in U$, the path from $v_r$ to one of these nodes does not pass through the other node. Next, let us consider the problem of maximizing
\bean
k'(U) & = & \sum_{u \in U } (\ell - L_T(u) ) 
\eean 
over all prefix-free subsets $U$ of $V$. It is straightforward to see that minimization of $\Bar{L}({\cal A})$ over all prefix-free codes is equivalent to maximisation of $k'(U)$ over all prefix-free $U\subset V$ of size $\ell$. Thus Huffman algorithm turns out to be an algorithm to identify a prefix-free set of nodes  
\bea \label{eq:maxp1}
U^* & \triangleq & \arg\max_{\substack{U \subset V, |U|=\ell \\ U \text{ is prefix-free}}} \sum_{u \in U} (\ell - L_T(u) ) 
\eea 
in a perfect binary tree of depth $\ell$. By the second assertion of Lemma~\ref{lem:Huff}, every node $u \in U^*$ is such that $L_T(u) =  \log_2\ell$ when $\ell$ is a power of $2$. Let us next consider the case when $\ell$ is not a power of $2$. Again by Lemma~\ref{lem:Huff}, $L_T(u) =  \lfloor \log_2\ell \rfloor$ or $L_T(u) =  \lceil \log_2 \ell \rceil$. Let $M = \{ u \in U^* \mid L_T(u)=\lfloor \log_2 \ell \rfloor \}$ and $m = |M|$. For every node $u \in U^* \setminus M$, $L_T(u)= \lceil \log_2 \ell \rceil$. Since the canopies of nodes in $U$ form a partition on the set of leaves of $T$, we count all the leaves of $T$ in two different ways to obtain:
\bean 
(\ell - m) \cdot 2^{\ell - \lceil \log_2 \ell \rceil} + m \cdot 2^{\ell - \lceil \log_2 \ell \rceil +1} &= & 2^{\ell}, \ \  \ \ 
\Rightarrow m \ = \ 2^{\lceil \log_2 \ell \rceil} - \ell. 
\eean 
We observe that $m =\mu$ that is defined as part of Defn.~\ref{defn:fl}. Thus the sequence $(\ell-L_T(u), u \in U^*)$ in non-decreasing order is given by:
\bea \label{eq:slstartail}
\begin{array}{cl}
\underbrace{\ell- \lceil \log_2 \ell \rceil, \ldots, \ell- \lceil \log_2 \ell \rceil}_{(\ell-m) \text{ terms}}, \underbrace{\ell- \lfloor \log_2 \ell \rfloor, \ldots, \ell- \lfloor \log_2 \ell \rfloor}_{m \text{ terms}}, & \text{$\ell$ is not a power of $2$,} \\
\ell- \log_2 \ell, \ldots, \ell-  \log_2 \ell ,\ell- \log_2 \ell, &  \text{$\ell$ is a power of $2$.}
\end{array} 
\eea 

In the third step, we consider a variant of the maximisation problem in \eqref{eq:maxp1} which aligns with our problem of identifying an anchor-decodable sequence with maximum $k(s_{\ell})$. Let us define
\bea \label{eq:maxp2}
U_1^* & \triangleq & \arg \max_{\substack{U \subset V, |U|=\ell \\ U \text{ is prefix-free}}} \left[ \sum_{u \in U} (\ell - L_T(u) )  - \max_{u\in U} (\ell - L_T(u) )\right].
\eea 
In order to establish an equivalence between finding $U_1^*$ and our desired anchor-decodable sequence, let us consider any prefix-free set $U = \{u_1,u_2,\ldots, u_\ell\}$ such that $L_T(u_1) \leq  L_T(u_2) \leq \cdots \leq L_T(u_{\ell})$. Then define 
\bean 
 s_{\ell} & = & \big(s_{\ell}(1),s_{\ell}(2),\ldots,s_{\ell}(\ell-1),s_{\ell}(\ell)\big)\\ 
&\triangleq & \big(\ell-L_T(u_{\ell}), \ell-L_T(u_{\ell-1}),\ldots, \ell-L_T(u_2),\ell\big), 
\eean 
where the first $(\ell-1)$ entries are determined by $U$. Since $U$ is prefix-free and $|U|=\ell$, $U$ defines a prefix-free code for a source with alphabet size $\ell$. By Lemma.~\ref{lem:Kraft} and the fact that $\ell-L_T(u_1)$ is maximum in the set $\{\ell-L_T(u), u \in U\}$, we have
\bean
2^{\ell} - \sum_{i=2}^{\ell} 2^{\ell - L_T(u_i)} \underset{\eqref{eq:Kraft}}{\geq} 2^{\ell - L_T(u_1)} \geq 2^{\ell - L_T(u_2)}.
\eean 
This means that $2^{\ell} - \sum_{i=1}^{\ell-1}2^{s_{\ell}(i)} \geq  2^{s_{\ell}(\ell-1)}$ and hence the non-decreasing sequence $s_{\ell}$ satisfies the first condition in Defn.~\ref{defn:udec}. In addition, we observe that
\bea 
k(s_{\ell}) \ = \ \sum_i s_{\ell}(i) & = & \ell +  \left[ \sum_{u \in U} (\ell - L_T(u) )  - \max_{u\in U} (\ell - L_T(u) )\right] 
\eea 
Therefore finding $U_1^*$ is equivalent to finding an $s_{\ell}$ that maximizes $k(s_{\ell})$ while satisfying the first condition of Defn.~\ref{defn:udec}.

In the fourth step, we argue that $U_1^*=U^*$. Suppose that $U_1^* = \{ u_1^*, u_2^*,\ldots, u_\ell^*\}$ such that $L_T(u_1^*) \leq  L_T(u_2^*) \leq \cdots \leq L_T(u_{\ell}^*)$ and $U^* = \{u_{h1}, u_{h2},\ldots, u_{h\ell}\}$ such that $L_T(u_{h1}) \leq  L_T(u_{h2}) \leq \cdots \leq L_T(u_{h\ell})$. At the outset, we clarify a subtle point with regard to the definition of both $U^*$ and $U_1^*$. If there are multiple candidates for $U_1^*$ or $U^*$, then we pick a common random element from the intersection of those candidate sets as the choice for both $U_1^*$ and $U^*$. Thus whenever there is a non-trivial intersection for these candidate sets, $U_1^*=U^*$. Suppose that $U_1^*\neq U^*$. This implies that there is no single $U$ that is a maximizer for both problems \eqref{eq:maxp1} and \eqref{eq:maxp2} simultaneously. Hence it must also be true that
\begin{equation} \label{eq:u1u}
 \Biggl[ \sum_{u \in U_1^*} (\ell - L_T(u) )  - \max_{u\in U_1^*} (\ell - L_T(u) )\Biggr] > \Biggl[ \sum_{u \in U^*} (\ell - L_T(u) )  - \max_{u\in U^*} (\ell - L_T(u) ) \Biggr].
\end{equation} 
Suppose that $\max_{u\in U_1^*} (\ell - L_T(u) ) < \max_{u\in U^*} (\ell - L_T(u) )$. By the second assertion in Lemma~\ref{lem:Huff}, $\ell-L_T(u_{hi})$ is either equal to or one less than $\max_{u\in U^*} (\ell - L_T(u) )$ for every $u_{hi} \in U^*$. This implies that $\max_{u\in U_1^*} (\ell - L_T(u) ) \leq \min_{u\in U^*} (\ell - L_T(u) )$ and therefore \eqref{eq:u1u} can not be true leading to a contradiction. So let us assume that $\max_{u\in U_1^*} (\ell - L_T(u) ) \geq \max_{u\in U^*} (\ell - L_T(u) )$. In that case, \eqref{eq:u1u} implies that 
\bean 
 \sum_{u \in U_1^*} (\ell - L_T(u) ) > \sum_{u \in U^*} (\ell - L_T(u) ). 
\eean 
This is a contradiction to the fact that $U^*$ is a maximizer for the problem in \eqref{eq:maxp1}. Hence we have proved that $U_1^* = U^*= \{ u_1^*, u_2^*,\ldots, u_\ell^*\}$.  Therefore, within the set of all sequences that is constrained by the first condition in Defn.~\ref{defn:udec}, the sequence
\bea
\nonumber s_{\ell}^* & = & \big(s^*_{\ell}(1),s^*_{\ell}(2),\ldots,s^*_{\ell}(\ell-1),s^*_{\ell}(\ell)\big)\\ 
 \label{eq:slstar}&\triangleq & \big(\ell-L_T(u^*_{\ell}), \ell-L_T(u^*_{\ell-1}),\ldots, \ell-L_T(u^*_2),\ell\big),  
\eea 
where $\ell-L_T(u_i^*)$ is as given in \eqref{eq:slstartail} in the same order, maximises $k(s_{\ell})$. 

As the final step to complete the proof, we proceed to check if $s_{\ell}^*$ satisfies the second condition in Defn.~\ref{defn:udec}. Let us define
\bean
\boldsymbol{\gamma}^* = (2^{\ell} - 1 - \sum_{i=1}^{\ell-1} 2^{s_{\ell}^*(i)}, \ 2^{s_{\ell}^*(i)}-1, i = \ell-1,\ldots, 1) .
\eean 
Let us consider the case when $\ell$ is not a power of $2$. Recall that $m$ is the number of $u\in U^*$ satisfying $L_T(u)=\lfloor \log_2 \ell \rfloor$. Then it can be computed that
\bean
\boldsymbol{\gamma}^* & = &      \Bigl(\underbrace{\frac{2^{\ell}}{2^{\lfloor \log_2 \ell \rfloor}} -1,\ldots, \frac{2^{\ell}}{2^{\lfloor \log_2 \ell \rfloor}} -1}_{m \text{ terms}}, \underbrace{\frac{2^{\ell}}{2^{\lceil \log_2 \ell \rceil}} -1, \ldots, \frac{2^{\ell}}{2^{\lceil \log_2 \ell \rceil}} -1}_{\ell-m \text{ terms}} \Bigr) .
\eean 
Since $1 \leq m \leq \ell - 2$, $\boldsymbol{\gamma}^*$ can not be equal to ${\sf cshift}(\boldsymbol{\gamma}^*,\ell_0)$ for any $1 \leq \ell_0 < \ell$. Since $s_{\ell}^* = f_{\ell}$ when $\ell$ is not a power of $2$, we have completed the proof for that case. 

What remains is the case when $\ell$ is a power of $2$. In this case, $\boldsymbol{\gamma}^* = (\frac{2^{\ell}}{\ell}-1, \frac{2^{\ell}}{\ell}-1, \ldots, \frac{2^{\ell}}{\ell}-1)$ and clearly ${\sf cshift}(\boldsymbol{\gamma}^*,\ell_0) = \boldsymbol{\gamma}^*$ for every $\ell_0$. Thus $s_{\ell}^*$ violates the second condition of Defn.~\ref{defn:udec} and therefore is not anchor-decodable. Since
\bean
k(s_{\ell}^*) & = & \ell^2 - \ell \log_2 \ell  + \log_2 \ell, 
\eean 
$\max k(s_{\ell}) \leq \ell^2 - \ell \log_2 \ell  + \log_2 \ell$ where the maximisation is over the set of all anchor-decodable sequences. On the other hand, we have
\bean 
k(f_{\ell}) & = &  \ell^2 - \ell \log_2 \ell  + \log_2 \ell  - 1  \ = \ k(s_{\ell}^*)-1, 
\eean
and $f_\ell$ is anchor-decodable. Therefore the optimality of $f_{\ell}$ follows if we prove that $k(s_{\ell}) \neq \ell^2 - \ell \log_2 \ell  + \log_2 \ell$ for any anchor-decodable  $s_{\ell}$. Suppose on the contrary $k(\hat{s}_{\ell})= \ell^2 - \ell \log_2 \ell  + \log_2 \ell$ for some anchor-decodable sequence $\hat{s}_{\ell}$. The vector of lengths $(\log_2\ell,\ell-\hat{s}_{\ell}(\ell-1),\ell-\hat{s}_{\ell}(\ell-2),\ldots, \ell-\hat{s}_{\ell}(1))$ has average length $\log_2\ell$, noting that $\hat{s}_\ell(\ell) = \ell$ due to the definition of an anchor-decodable sequence. Since $\hat{s}_{\ell}$ respects the first condition of Defn.~\ref{defn:udec}, we must have
\bea
\nonumber 2^{\ell} - \sum_{i=1}^{\ell-1} 2^{\hat{s}_{\ell}(i)} & \geq & 2^{\hat{s}_{\ell}(\ell-1)} = 2^{\max_{i=1,\ldots,\ell-1}\hat{s}_{\ell}(i)} \\
\nonumber \Rightarrow 1 - \sum_{i=1}^{\ell-1} 2^{-(\ell-\hat{s}_{\ell}(i))} & \geq & 2^{- \min_{i=1,\ldots,\ell-1} (\ell- \hat{s}_{\ell}(i))}\\
\label{eq:Kraftlen} \Rightarrow \sum_{i=1}^{\ell-1} 2^{-(\ell-\hat{s}_{\ell}(i))} + 2^{-\log_2\ell} & \leq & 1  .
\eea 
The inequality in \eqref{eq:Kraftlen} is true because the average of $(\ell-1)$ numbers $\ell-\hat{s}_{\ell}(\ell-1),\ell-\hat{s}_{\ell}(\ell-2),\ldots, \ell-\hat{s}_{\ell}(1)$ can be computed as $\log_2\ell$ and hence $\min_{i=1,\ldots,\ell-1} (\ell- \hat{s}_{\ell}(i)) \leq \log_2\ell$. By \eqref{eq:Kraftlen} and Lemma~\ref{lem:Kraft}, the length vector $(\log_2\ell,\ell-\hat{s}_{\ell}(\ell-1),\ell-\hat{s}_{\ell}(\ell-2),\ldots, \ell-\hat{s}_{\ell}(1))$ corresponds to a prefix-free code of average length $\log_2\ell$. If $T_p$ is the code tree of the code, then by the third statement of Lemma~\ref{lem:Huff}, $L_{T_p}(v_i) = \log_2\ell$ for every $i$. Therefore $\hat{s}_{\ell}(i)=\ell-\log_2\ell$ for every $i=1,2,\ldots,\ell-1$. Thus $\hat{s}_\ell$ becomes equal to $s_{\ell}^*$ leading to a contradiction to the assumption that $\hat{s}_{\ell}$ is anchor-decodable. It follows that $k(s_\ell)$ is maximized by choosing $s_{\ell} = f_{\ell}$ when $\ell$ is a power of $2$. This completes the proof.
\epf 
\section{A Second Code Construction \label{sec:second}}

As established by Theorem~\ref{thm:cloptimal}, the code ${\cal C}[\ell]$ has the maximum combinatorial dimension among the family of codes $\{{\cal C}[s_\ell] \mid s_{\ell} \text{ is anchor-decodable}\}$ that are encoded by Algorithm~\ref{alg:g1} and are decodable by Algorithm~\ref{alg:g2}. Both these algorithms are of very low complexity. Two questions that arise at this point are: 
\ben
\item Can Algorithm~\ref{alg:g1} generate fast decodable codes when $s_{\ell}$ is not necessarily anchor-decodable? 
\item Is $f_{\ell}$ a unique sequence and hence ${\cal C}[\ell]$ a unique code that achieves the maximum combinatorial dimension $k_{\ell}$?
\een 
In this section, we answer the first question in the affirmative by a presenting a code generated by Algorithm~\ref{alg:g1} picking as auxiliary input a sequence that is not anchor-decodable, yet admitting an alternate decoder that has the same order of complexity as that of Algorithm~\ref{alg:g2}. As evident in the proof of Theorem~\ref{thm:code}, the crux of the decoding algorithm in Alg.~\ref{alg:g2} lies in the fact that the maximum gap between two successive $1$'s in a codeword of ${\cal C}[s_{\ell}]$ uniquely identifies the anchor when $s_{\ell}$ is anchor-decodable. It turns out that we can come up with an alternate fast decoding algorithm that relies not just on the maximum gap, but on a subset of gaps containing the maximum one. Interestingly, the combinatorial dimension of such a new code matches with $k_{\ell}$ for certain values of $\ell$ and thereby, establishing that the sequence $f_{\ell}$ is not unique in that sense. This answers the second question in the negative. 

\subsection{An Alternate Integer Sequence \label{subsec:seq2}}

Let $\ell$ and $r$ be two integer parameters satisfying $\ell \geq 3$ and $1 \leq r \leq \lfloor \frac{\ell+3}{4} \rfloor$. In this subsection, we define a sequence $f_{\ell,r}$ that is not anchor-decodable. 
\bdefn \label{defn:flr} Let $\ell \geq 3$ and $1 \leq r \leq \lfloor \frac{\ell+3}{4} \rfloor$. Then 
\bea \label{eq:flr1}
f_{\ell, r}(i) = \left\{ \begin{array}{ll}
        r + \delta(\ell,r) & \ \ i =1 \\ 
        r + i -1, &  \ \ i = 2, 3, \ldots, \ell-2r-1 \\ 
        \ell - 1 - \lceil \frac{\ell-i}{2} \rceil,   & \ \ i = \ell-2r, \ell-2r+1, \ldots, \ell-1\\
        \ell,   & \ \ i = \ell
\end{array} \right. 
\eea 
where 
\bea \label{eq:flr2}
\delta(\ell, r) = \left\{ \begin{array}{ll}
        1,   & \ \ \ell > 2r+2 \\
        0,     & \ \ (r = 1, \ell \in \{3,4\}) \text{ or } (r = 2, \ell \in \{5,6\}) 
\end{array} \right. 
\eea 
\edefn 
Observe that $\delta(\ell,r)$ equals $1$ for every permitted value of $\ell,r$ except for a limited set of parameters $(r=1,\ell=3,4)$ and $(r=2,\ell=5,6)$. Next we define
\bea 
\label{eq:klr} k_{\ell, r} & \triangleq & \sum_{i=1}^\ell f_{\ell, r}(i) \ = \
    \frac{\ell(\ell - 1)}{2} + r(\ell-r-1) + 1 + \delta(\ell,r) . 
\eea
\setlength{\tabcolsep}{5pt}
\begin{table}[ht]
	\centering 
		\begin{tabular}{|c|c|c|c|c|} 
			\hline 
                $\ell$ & $f_{\ell}$ & $\hat{f}_{\ell}$ & $k_{\ell}$ & $\hat{k}_{\ell}$ \\
                \hline \hline  
                 3 & 1, 1, 3 & 1, 1, 3 & 5 & 5 \\
                \hline
                 4 & 1, 2, 2, 4 & 1, 2, 2, 4 & 9 & 9 \\
                \hline
                 5 & 2, 2, 3, 3, 5 & 2, 2, 3, 3, 5 & 15 & 15 \\
                \hline
                 6 & 3, 3, 3, 3, 4, 6 & 2, 3, 3, 4, 4, 6 & 22 & 22 \\
                \hline
                 7 & 4, 4, 4, 4, 4, 4, 7 & 3, 3, 4, 4, 5, 5, 7 & 31 & 31 \\
                \hline
                 8 & 4, 5, 5, 5, 5, 5, 5, 8 & 3, 3, 4, 5, 5, 6, 6, 8 & 42 & 40 \\
                \hline
                 9 & 5, 5, 6, 6, 6, 6, 6, 6, 9 & 4, 4, 5, 5, 6, 6, 7, 7, 9 & 55 & 53 \\
                \hline
                 10 & 6, 6, 6, 6, 7, 7, 7, 7, 7, 10 & 4, 4, 5, 6, 6, 7, 7, 8, 8, 10 & 69 & 65 \\
                \hline                
    \end{tabular} 
    \vspace{0.1in}
	\caption{\centering Compilation of $f_{\ell}$ and $\hat{f}_{\ell}$.   \label{tab:f}}
\end{table} 
We compile certain useful numerical identities pertaining to the sequence in the following proposition.

\bprop \label{prop:flr} The following identities hold:
\ben 
\item $f_{\ell, r}(i) > \ell - r -1, \ \ i = \ell-1,\ell-2,\ldots, \ell-2r+2$.
\item $f_{\ell, r}(i) = \ell - r -1, \ \ i = \ell-2r+1,\ell-2r$.
\item $f_{\ell, r}(i) < \ell - r -1, \ \ i = \ell-2r-1,\ell-2r-2,\ldots, 1$.
\item When $r_1 < r_2 \leq \left\lfloor \frac{\ell + 3}{4} \right\rfloor$, $f_{\ell r_1}(i) \leq f_{\ell r_2}(i)$ for every $i \in [\ell]$. 
\item When $r_1 < r_2 \leq \left\lfloor \frac{\ell + 3}{4} \right\rfloor$, $k_{\ell, r_1} < k_{\ell, r_2}$.
\een 
\eprop 
\bpf They all follow from definitions in a straightforward manner. It is necessary to have $\delta(\ell,r)=0$ when $\ell \leq 2r+2$ for the first three identities to hold.
\epf
By fifth property of Prop.~\ref{prop:flr}, $k_{\ell r}$ is maximized at $r_{\text{max}} = \left\lfloor \frac{\ell + 3}{4} \right\rfloor$, and we define $\hat{f}_{\ell} = f_{\ell, r_{\text{max}}}$. We also define 
\bea \label{eq:klhat}
\hat{k}_{\ell} & \triangleq & \sum_i \hat{f}_{\ell}(i) \ = \ \max_r k_{\ell, r} \ = \ \frac{\ell(\ell-1)}{2} + \left\lfloor \frac{\ell + 3}{4} \right\rfloor \left( \left\lceil \frac{3(\ell - 1)}{4} \right\rceil - 1 \right) + \delta_{\ell} 
\eea
where 
\bea
\delta_{\ell} & = & \left\{ \begin{array}{ll}
     1, & \ \ \ \ell > 6  \\
     0, & \ \ \ 1 \leq \ell \leq 6 
\end{array} \right. .
\eea 
A compilation of $\hat{f}_{\ell}$ and $f_{\ell}$ along with corresponding values of $\hat{k}_{\ell}$ and $k_{\ell}$ is provided in Table~\ref{tab:f}.

\subsection{The Encoding Algorithm}
\label{subsec:encoding2}

As described in Sec.~\ref{subsec:encoding}, Algorithm~\ref{alg:g1} provides a generic encoding method because it can be invoked with any $\ell$-length auxiliary input sequence $s_{\ell}$ such that $s_{\ell}(\ell)=\ell$. We invoke it with the choice $s_{\ell} = f_{\ell,r}$. In the following Lemma~\ref{lem:wt2}, we show that the Hamming weight of the encoder output is always $\ell$ despite the fact that $f_{\ell, r}$ is not anchor-decodable.

\blem \label{lem:wt2} With $s_{\ell}= f_{\ell, r}$ as defined in Definition~\ref{defn:flr}, $w_H(\phi({\bf x})) = \ell$ for every ${\bf x} \in \{0,1\}^{k_{\ell r}}$ where $\phi(\cdot)$ is determined by Alg.~\ref{alg:g1}. \elem 
\bpf We follow the same line of arguments as in the proof of Lem.~\ref{lem:wt}. The maximum cumulative increment $p$ in the variable ${\sf pos}$ over the last $(\ell-1)$ iterations of the loop in \textsl{Line} $4$ is given by:
\bea 
\nonumber p & = & \sum_{j=1}^{\ell-1} 2^{|{\bf x}_j|}  \ = \ \sum_{j=1}^{\ell-1} 2^{f_{\ell, r}(j)}  \\ 
\nonumber & = & \sum_{j=2}^{\ell-2r-1} 2^{j+r-1} + \sum_{j=\ell-2r}^{\ell-1} 2^{\ell-1-\lfloor \frac{\ell-j}{2}\rfloor} + 2^{r+\delta(\ell,r)} \\
\nonumber & = & \sum_{j=1}^{\ell-2r-1} 2^{j+r-1} + \sum_{j=\ell-2r}^{\ell-1} 2^{\ell-1-\lfloor \frac{\ell-j}{2}\rfloor} + \delta(\ell,r) 2^{r}  \\
\label{eq:p3} & = & \left\{ \begin{array}{ll} 2^{\ell} - 2^{\ell-r-1} - 2^{r}, & \ \ell > 2r+2 \\
2^{\ell} - 2^{\ell-r-1}, & \ \ell \leq  2r+2 
\end{array} \right. 
\eea  
Since $p < 2^{\ell}$ by \eqref{eq:p3}, a distinct bit of ${\bf c}$ is set from $0$ to $1$ in each of these $(\ell-1)$ iterations and therefore $w_H({\bf c})=\ell$.
\epf


\subsection{A Decoding Algorithm and a Constant Weight Code}
\label{subsec:decoding2}


Let ${\bf c}$ be an output of the encoder. In order to decode the input ${\bf x}$ uniquely, it is necessary and sufficient to identify the anchor. However, the sequence $f_{\ell, r}$ is not anchor-decodable, and therefore the procedure \textsc{FindAnchor} in Alg.~\ref{alg:g3} will not work. Nevertheless, we illustrate with an example of $\ell=7, r=2$ that it is possible to determine the anchor bit based on the pattern of gaps in ${\bf c}$ (See Fig.~\ref{fig:decode} for a pictorial illustration.). Continuing the approach taken in the description of warm-up construction (see Fig.~\ref{fig:warm-up} in Sec.~\ref{subsec:warm-up}), the codeword of length $n=128$ is represented as a circle with $128$ points indexed from $0$ to $127$. The codeword ${\bf c}$ picked in the example has $c[j]=1$ for $j =10, 26, 32, 37, 64, 96, 127$ and zero everywhere else. To avoid clutter in Fig.~\ref{fig:decode}, we indicate the starting point $0$ and mark only those points at which $c[j]=1$, instead of all the $128$ points.

First, we identify the gaps between successive $1$'s as ${\bf g}[m], m=0,1,\ldots, 6$ in order starting from the first gap ${\bf g}[0] = {\sf gap}(127,10) = 10$. Other gaps are ${\bf g}[1] = 15, {\bf g}[2] = 5, {\bf g}[3]= 4, {\bf g}[4]= 26, {\bf g}[5]= 31, {\bf g}[6]= 30$. The principle is to look for a stretch of $(2r-1)=3$ consecutive gaps in clockwise direction such that the last gap in each of these stretch is $\geq 2^{\ell-r-1} = 16$. The gap that is on or above the threshold $2^{\ell-r-1}$ is referred to as a {\em candidate gap}. There are three such stretches marked in this example, marked as $\textcircled{a}$, $\textcircled{b}$ and $\textcircled{c}$ in Fig.~\ref{fig:decode}. Among these three, the stretch $\textcircled{c}$ containing $({\bf g}[2], {\bf g}[3], {\bf g}[4])= (5,4,26)$ is unique in the sense that every gap in that stretch apart from the last gap ${\bf g}[4]$ does not qualify as a candidate gap. The bit $c[64]$ at the end of $\textcircled{c}$ is therefore picked as the anchor bit. Once the anchor is identified as $c[64]$, binary equivalent of $64$ gives rise to ${\bf x}_{7}$, and that of following six gaps $({\bf g}[5], {\bf g}[6],{\bf g}[0],{\bf g}[1], {\bf g}[2], {\bf g}[3])= (31, 30, 10, 15, 5, 4)$ yield ${\bf x}_{6},{\bf x}_{5},{\bf x}_{4},{\bf x}_{3},{\bf x}_{2}$ and ${\bf x}_{1}$. Except when $\delta(\ell, r)=1$ and a specific type of message vector appears, the above procedure for finding the anchor bit works. The correctness of the above procedure and the way to handle special cases constitute the following theorem. 
 
\begin{figure}[htb!]
    \centering
    \includegraphics[scale=0.18]{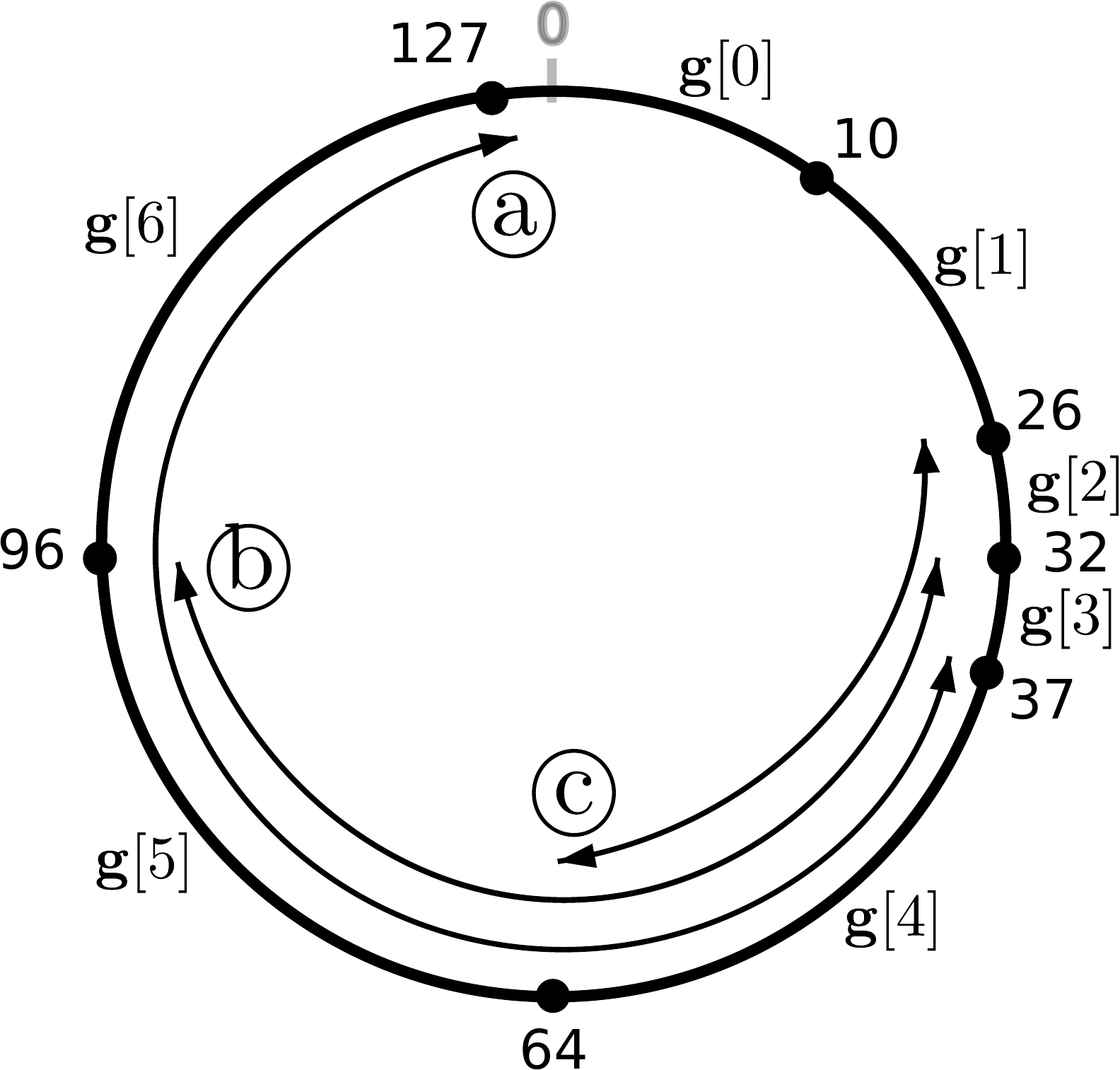}
    \caption{Illustration of the principle of decoding algorithm for  $\ell=7, r=2$ when the codeword ${\bf c}$ has $1$'s at $c[j], j =10, 26, 32, 37, 64, 96, 127$ (marked with dots) and $0$'s everywhere else. There are three clock-wise stretches of gaps marked as $\textcircled{a}$, $\textcircled{b}$ and $\textcircled{c}$ that end in a candidate gap, i.e., with value  on or above $16$. The stretch $\textcircled{c}$ given by $(5,4,26)$ is unique among these three because in $\textcircled{c}$, every gap value apart from the last one does not qualify as a candidate. The bit $c[64]$ at the end of the stretch $\textcircled{c}$ is therefore picked as the anchor bit.}
    \label{fig:decode}
\end{figure}

\bthm \label{thm:code2} When the auxiliary input is chosen as $f_{\ell, r}$ given in Definition~\ref{defn:flr}, the map $\phi$ defined by Algorithm~\ref{alg:g1} is one-to-one.
\ethm 
\bpf 
\begingroup
\begin{algorithm}
	\caption{\textsc{Decode2}  \newline {\bf Input}: ${\bf c} \in \text{Im}(\phi), s_{\ell} $ \newline {\bf Output}: ${\bf x} \in \{0,1\}^{k(s_{\ell})}$ \label{alg:g4}}
	\DontPrintSemicolon
	Find $0 \leq j[0] < j[1] < \cdots < j[\ell -1 ] < n$ such that $c[j[i]] = 1$ for every $i=0,1,\ldots, \ell-1$. 
    
    $\gaps[m] = {\sf gap}(j[(m-1) \mod \ell], j[m])$ for $m=0,1,\ldots \ell-1$

    ${\sf anchor\_index} = \textsc{FindAnchor2}({\bf g}, s_{\ell})$
    
    Initialize binary vector ${\bf x}$ such that $|{\bf x}|=\ell$ and $\text{dec}({\bf x})=j[{\sf anchor\_index}]$
    
    \For{$i = 1,2,\ldots, \ell-1$}    
    {
		$g \leftarrow \gaps[({\sf anchor\_index}+i) \mod \ell]$
		
		Represent $g$ as binary string ${\bf x}_{i}$ of length $s_{\ell}(\ell-i)$

		${\bf x} \leftarrow {\bf x}\Vert {\bf x}_{i}$
    }
\end{algorithm}
\endgroup

\begingroup
\begin{algorithm}
	\caption{\textsc{FindAnchor2}  \newline {\bf Input}: ${\bf g} \in \mathbb{Z}_n^{\ell}, f_{\ell, r}$ \newline {\bf Output}: ${\sf anchor\_index} \in [0 \ \ell-1]$ \label{alg:g5}}
	\DontPrintSemicolon
    ${\sf gaps\_allone} \leftarrow (2^{\ell-r-1}-1) \Vert (2^{f_{\ell, r}(i)}-1,i=\ell-1,\ell-2,\ldots, 1)$ 
    
    \If{ $\delta(\ell, r)=1$ {\em and} $\exists n_0 \in \mathbb{Z}_{\ell}$ \emph{such that} ${\sf gaps\_allone} = {\sf cshift}({\bf g},n_0)$ }
    {
        ${\sf anchor\_index} \leftarrow n_0$
        
    } \Else
    {
        Initialize array ${\sf cadidates}$ to $0^{\ell}$.
    
        \For{$m = 0,1,\ldots, \ell-1$}
        {
            \If{$\gaps[m] \geq 2^{\ell-r-1}$}
            {
                ${\sf cadidates}[m] \leftarrow 1$ 
            }
        }
        Pick $m_0$ such that ${\sf cadidates}[m_0] = 1$ 
        
        ${\sf anchor\_index} \leftarrow m_0$
        
        ${\sf non\_cand\_cnt\_bkwd} \leftarrow 0$
    
        \For{$m = (m_0+1 \mod {\ell}) ,(m_0+2 \mod {\ell}), \ldots, (m_0+\ell \mod {\ell})$}
        {
            \If{${\sf cadidates}[m] = 0$}
            {
                ${\sf non\_cand\_cnt\_bkwd} \leftarrow {\sf non\_cand\_cnt\_bkwd} + 1$
            
            } \Else 
            {
                \If{${\sf non\_cand\_cnt\_bkwd} \geq 2r-2$}
                {
                    ${\sf anchor\_index} \leftarrow m$
            
                    \textbf{break}
                }    
                ${\sf non\_cand\_cnt\_bkwd} \leftarrow 0$
            }
        }
    }
\end{algorithm}
\endgroup

Let ${\bf c}$ be an arbitrary output of the encoder when the input ${\bf x} = {\bf x}_{\ell} \Vert {\bf x}_{\ell-1} \Vert \cdots \Vert {\bf x}_{1}  \in \{0,1\}^{k_{\ell r}}$ where $|{\bf x}_i|=f_{\ell,r}(i)$.  Along the lines of the proof of Theorem~\ref{thm:code}, we provide a explicit decoder for ${\bf c}$ that maps uniquely to ${\bf x}$. The decoder as given in Algorithm~\ref{alg:g4} is exactly the same in Alg.~\ref{alg:g3} except for the fact that {\sf anchor\_index} is determined by invoking a different procedure \textsc{FindAnchor2} presented in Algorithm~\ref{alg:g5}. The part of the proof that argues correctness of Algorithm~\ref{alg:g4} once {\sf anchor\_index} is correctly determined remains the same as that of Theorem~\ref{thm:code} and we do not repeat it here. The notations $j[m], m=1,2,\ldots, \ell-1, j[{\sf anchor\_index}]$ and $\gaps \in \mathbb{Z}_n^{\ell}$ also remain the same. 

It is sufficient to argue that the procedure \textsc{FindAnchor2} is correct. Following the same line of arguments after \eqref{eq:gapsum}, we have 
\bea 
\nonumber \gaps[{\sf anchor\_index}]  & = & (n - \ell) - \sum_{i = 1}^{\ell-1} \gaps[({\sf anchor\_index}+i )  \hspace{-2.5mm}\mod \ell] \\
 \label{eq:initlb1} & \geq & (n - \ell) - \sum_{i = 1}^{\ell-1} (2^{|{\bf x}_{\ell-i}|} - 1) \\ 
 \nonumber & = & (2^{\ell} - \ell) - \sum_{i = 1}^{\ell-1} (2^{f_{\ell, r}(\ell -i)} - 1) \\
\label{eq:initlb2} & = & \left\{ \begin{array}{ll}
     2^{\ell-r-1} - 1, &  \ \delta(\ell,r) = 1 \\
     2^{\ell-r-1} + 2^r -1, & \ \delta(\ell,r) = 0
\end{array} \right. 
\eea 
The inequality in \eqref{eq:initlb1} follows from the way ${\bf x}_{\ell-i}$ is encoded by Algorithm~\ref{alg:g1}. It is straightforward to check that equality holds in \eqref{eq:initlb1} if and only if the message vector is of the type
\bea \label{eq:allone}
{\bf x}_{\ell - i} & = & 1^{f_{\ell, r}(\ell - i)}, \text{~~for all}~ i=1,2,\ldots, \ell-1. 
\eea 
When the message vector satisfies \eqref{eq:allone}, every gap except $\gaps[{\sf anchor\_index}]$ becomes maximal in length, and therefore we refer to this special case as the {\em maximal-gap} case. When $\delta(\ell,r)=1$, \textsl{Lines} $2-3$ in Alg.~\ref{alg:g5} checks for the maximal-gap case by comparing every circular shift of the vector ${\bf g}$ with a fixed vector ${\sf gaps\_allone}$. The vector ${\sf gaps\_allone}$ corresponds to a message vector of the type
\bea \label{eq:gapallone2}
{\bf x}_{\ell - i} & = & 1^{f_{\ell, r}(\ell - i)},\text{~for all}~ i=1,2,\ldots, \ell-1, \text{~~and}~~  \text{dec}({\bf x}_{\ell}) \ \leq \ 2^{\ell-r-1}-1 . 
\eea 
for which ${\sf anchor\_index} = 0$. If ${\sf cshift}({\bf g},n_0)$ becomes equal ${\sf gaps\_allone}$ for some $0 \leq n_0 \leq (\ell-1)$, then by the first three identities of Prop.~\ref{prop:flr}, $n_0$ is unique and is equal to ${\sf anchor\_index}$. 

If  \eqref{eq:allone} is false, then clearly \eqref{eq:initlb1} satisfies with strict inequality, and in that case $\gaps[{\sf anchor\_index}] \geq 2^{\ell-r-1}$ for every $\ell,r$ by \eqref{eq:initlb2}. The binary array ${\sf cadidates}$ generated after the execution of the loop in \textsl{Line} $8$ is such that ${\sf cadidates}[m]=1, m \in [0 \ \ell-1]$ if and only if $\gaps[m] \geq 2^{\ell-r-1}$, and therefore the binary array  ${\sf cadidates}$ keeps a record of all gaps that can possibly be a candidate for ${\bf g}[{\sf anchor\_index}]$. As already made clear, ${\sf anchor\_index}$ is indeed picked as a candidate. As a result, it becomes possible to execute \textsl{Line} $9$ as there is always an $m_0$ such that ${\sf cadidates}[m_0] =1$. If there are no other candidates, ${\sf anchor\_index}$ is indeed $m_0$. This is exactly what the procedure returns as the value of ${\sf anchor\_index}$ is not changed after executing \textsl{Line} $10$.

Let us investigate how the procedure works when there are more than one candidates for ${\sf anchor\_index}$. If $r=1$, we observe that 
\bean 
2^{\ell-r-1}  =  2^{\ell-2}  & > & \max_{i \in [\ell-1]} \ (2^{f_{\ell, r}(i)} - 1) \\
& \geq & \gaps[m], \ m \neq {\sf anchor\_index}
\eean 
and therefore there shall be exactly one candidate for ${\sf anchor\_index}$ and we fall back to the previous case. So in the discussion on having multiple candidates, we assume that $r \geq 2$. By the second and third identities in Prop.~\ref{prop:flr}, we have
\bea
\nonumber 2^{\ell - r - 1} & > & (2^{f_{\ell, r}(i)} - 1), \ i = \ell-2r+1, \ell - 2r, \ldots,1  \\
\nonumber & = & (2^{|{\bf x}_{i}|} - 1) \\
\label{eq:behindinit1} & \geq & \gaps[({\sf anchor\_index}  - i ) \mod \ell ] , \ i = \ell-2r+1, \ell-2r, \ldots,  2,1.
\eea 
Since $\ell \geq 4r-3$, we have $\ell-2r+1 \geq 2r-2$. In addition, since $\ell \geq 3$ and $\ell \geq 4r-3$, we have $\ell > 2r$. Therefore, the set $\{\ell-2r+1,\ell-2r, \ldots, 1\}$ contains the subset $\{1,2,\ldots, 2r-2\}$ which is non-empty as $r \geq 2$. Thus \eqref{eq:behindinit1} implies a non-vacuous statement 
\bea \label{eq:behindinit2}
\gaps[({\sf anchor\_index}  - i )  \hspace{-2.5mm}\mod \ell ] & < &  2^{\ell-r-1}, \ i = 1, 2,\ldots, 2r - 2.
\eea 
The loop at \textsl{Line} $12$ begins its iterations starting with $m=(m_0+1) \mod \ell$ where $m_0$ corresponds to a candidate gap. As a consequence, in every iteration of the loop indexed by $m$, the variable ${\sf non\_cand\_cnt\_bkwd}$ acts as a counter for the number of gaps to the left of $\gaps[m]$ (counting cyclically) that do not qualify as candidates until a candidate is met. It follows from \eqref{eq:behindinit2} that when $m={\sf anchor\_index}$, ${\sf non\_cand\_cnt\_bkwd} \geq 2r-2$, and therefore \textsl{Lines} $17-18$ get executed if the loop prolongs enough to witness $m={\sf anchor\_index}$. Suppose $m' \neq {\sf anchor\_index}$ corresponds to a candidate gap, i.e., ${\sf cadidates}[m'] =1$, then it means that $\gaps[m'] \geq  2^{\ell - r - 1}$. But we know that $m' = ({\sf anchor\_index} + i') \mod \ell$ for some $i' \in [\ell-1]$ and $\gaps[m']  < 2^{f_{\ell, r}(\ell-i')}$. Since $m'$ is a candidate, $i'$ must satisfy that 
\bean 
2^{\ell - r - 1} & <  & 2^{f_{\ell, r}(\ell-i')}  
\eean 
and it follows from the first identity in Prop.~\ref{prop:flr} that
\bea
\nonumber  i' & \leq & 2r-2 \\ 
\label{eq:frontinit} \Rightarrow m' & \leq  & {\sf anchor\_index} + (2r-2) \mod \ell.
\eea 
It follows from \eqref{eq:frontinit} that for any candidate $m' \neq {\sf anchor\_index}$, the number of non-candidate gaps to the left of $m'$ (cyclically) is strictly less than $2r-2$. In other words, it must be that ${\sf non\_cand\_cnt\_bkwd} < 2r-2$, and therefore \textsl{Lines} $17-18$ will not get executed for $m'$. Therefore the iterations of the loop until $m = {\sf anchor\_index}$ happens. Thus we have shown that the \textsl{Lines} $17-18$ get executed if and only if $m = {\sf anchor\_index}$. Therefore the procedure \textsc{FindAnchor2} to determine {\sf anchor\_index} uniquely is indeed correct when there are multiple candidates. This completes the proof.
\epf 
The decoding algorithm as presented in Algorithm~\ref{alg:g4} and Algorithm~\ref{alg:g5}  illustrates the principle of operation but can be implemented as a single-pass loop on $n$ bits using a circular buffer. Therefore it has the same order of complexity as that of Alg.~\ref{alg:g2}. By Lemma~\ref{lem:wt2} and Theorem~\ref{thm:code2}, ${\cal C}[f_{\ell,r}]$ is a binary constant weight code even if $f_{\ell,r}$ is not anchor-decodable. The sequence $\hat{f}_{\ell}$ obtained by specializing $f_{\ell,r}$ with $r=r_{\max}$ produces a code with maximum combinatorial dimension among $\{ {\cal C}[f_{\ell,r}] \mid 1 \leq r \leq r_{\max} \}$  leading to the following definition. 
\bdefn \label{defn:hatclcode} Let $\ell \geq 3$. We define the code ${\cal \hat{C}}[\ell] = {\cal C}[\hat{f}_{\ell}]$. The code ${\cal \hat{C}}[\ell]$ has blocklength $n=2^\ell$, weight $w=\ell$, and combinatorial dimension $k={\hat{k}_{\ell}}$ as defined in \eqref{eq:klhat}.
\edefn  
\section{Properties of the Codes\label{sec:props}}

\subsection{On Codebook Size}
\label{subsec:code_size}

The following straightforward lemma gives an information-theoretic upper bound on the size of any binary constant weight code.

\blem \label{lem:ub} Let ${\cal C}$ be a constant weight binary code of blocklength $n$, weight $w$ and combinatorial dimension $k$. Then 
\bea \label{eq:ubi}
k & \leq & \lfloor \log_2 A(n,2,w) \rfloor \ = \ \left\lfloor \log_2 {n \choose w} \right \rfloor . 
\eea
\elem 
It is easy to see that both ${\cal C}[\ell]$ and ${\cal \hat{C}}[\ell]$ has minimum distance $d=2$ because $1^\ell\Vert 0^{n-\ell}$ and $0\Vert 1^{\ell}\Vert 0^{n-\ell-1}$ that are apart by Hamming distance $2$ are codewords in both the codes. Therefore, it is meaningful to compare their combinatorial dimensions against the bound in \eqref{eq:ubi}. If we substitute $n=2^{\ell}, w = \ell$ in \eqref{eq:ubi}, we obtain
\bea
\nonumber k & \leq & \left\lfloor \log_2 {2^{\ell} \choose \ell} \right \rfloor  \\ 
\label{eq:stirling} & \leq &  \log_2 \left[ \frac{1}{\sqrt{2\pi\ell}} \left( \frac{2^{\ell} e}{\ell}\right)^\ell \right] \\
\label{eq:ubi1} & = & \ell^2 - \ell \log_2 \ell + \tfrac{1}{\ln 2} \ell  - (\tfrac{1}{2})\log_2 \ell - \tfrac{\ln (2\pi)}{\ln 2}
\eea 
The inequality \eqref{eq:stirling} follows from Stirling's approximation. Along with \eqref{eq:ubi1}, another upper bound can be obtained owing to certain cyclic structure that our construction brings along.
\blem \label{lem:cshift} If ${\bf c} \in {\cal C}[\ell]$ (or ${\cal \hat{C}}[\ell]$), then ${\sf cshift}({\bf c},n_0) \in {\cal C}[\ell]$ (or ${\cal \hat{C}}[\ell]$) for every $n_0$. When $n_0 \in \mathbb{Z}_{2^\ell}$, ${\sf cshift}({\bf c},n_0)$ must be distinct for every distinct $n_0$.
\elem
\bpf Let ${\bf x} = \phi^{-1}({\bf c}) = {\bf x}_{\ell} \Vert {\bf \hat{x}}_{\ell}$ where $|{\bf x}_{\ell}|=\ell$ and $\phi$ is invoked with auxiliary input $f_{\ell}$ or $\hat{f}_{\ell}$ as the case may be. For every $n_0$, ${\sf cshift}({\bf c},n_0)$ is a codeword corresponding to a message obtained by updating ${\bf x}_{\ell}$ (if required), but keeping ${\bf \hat{x}}_{\ell}$ fixed. This proves the first claim. Consider all codewords obtained by varying ${\bf x}_{\ell}$ over all $2^\ell$ possibilities, but keeping ${\bf \hat{x}}_{\ell}$ fixed. They must all be distinct from one another because $\phi$ is one-to-one. Since there can at most be $2^{\ell}$ distinct cyclic shifts possible for ${\bf c}$, ${\sf cshift}({\bf c},n_0)$ must all be distinct for every  $n_0 \in \mathbb{Z}_{2^\ell}$.
\epf 
Let $C_n$ denote the cyclic group of order $n$. By Lemma~\ref{lem:cshift}, the action of $C_{2^\ell}$ on ${\cal C}[\ell, r]$ results in orbits of size $2^\ell$. This implies that ${\cal C}[\ell, r]/C_{2^\ell}$ contains only primitive binary necklaces of length $2^\ell$ and weight $\ell$. Recall that a binary necklace of length $n$ is an equivalence class of vectors in $\{0,1\}^n$ considering all the $n$ rotations of a vector as equivalent. A binary necklace is said to be primitive if the size of the equivalence class is $n$. The count of primitive binary necklaces of length $n$ and weight $w$ is known to be \cite{Riordan} 
\bea \label{eq:pnw}
p(n,w) & = & \sum_{d \mid n} \mu\left( \frac{n}{d} \right) q(d, wd/n)
\eea
where $q(d, wd/n)$ is the coefficient of $x^{wd/n}y^{(n-w)d/n}$ in the polynomial 
\bea \label{eq:Fxy}
F(x,y) & = & \frac{1}{n} \sum_{d \mid n} (x^{n/d} + y^{n/d})^d \phi_E\left( \frac{n}{d} \right) .
\eea 
Here $\mu(\cdot)$ and $\phi_E(\cdot)$ are M\"{o}bius function and Euler's totient function respectively. By Lemma~\ref{lem:cshift} and \eqref{eq:pnw}, both $k_{\ell}$ and $\hat{k}_{\ell}$ are upper bounded by 
\bea \label{eq:ubp}
 \lfloor \log_2 (np(n,w)) \rfloor \ = \ \ell + \lfloor \log_2 p(2^\ell, \ell)  \rfloor .
\eea
It is not clear when the bound in \eqref{eq:ubp} is strictly better than the one in \eqref{eq:ubi} for an arbitrary value of $\ell$. In any case, the sizes of both ${\cal \hat{C}}[\ell]$ and ${\cal C}[\ell]$ must respect both the upper bounds \eqref{eq:ubi1} and \eqref{eq:ubp}. 

Comparing the lower bound in Prop~\ref{prop:klbound} and the upper bound in \eqref{eq:ubi1}, it is worthwhile to make the following inferences on the performance of ${\cal C}[\ell]$ and ${\cal \hat{C}}[\ell]$. When $\ell=3$, ${\cal \hat{C}}[3] = {\cal C}[3]$ and the code is optimal as $k_3=5$ matches the information-theoretic upper bound. The code ${\cal C}[4]$ (same as ${\cal \hat{C}}[4]$) has $k_4=9$ that is one bit away from the bound. While both ${\cal C}[\ell]$ and ${\cal \hat{C}}[\ell]$ have the same combinatorial dimension for $3 \leq \ell \leq 7$, ${\cal C}[\ell]$ clearly outperforms ${\cal \hat{C}}[\ell]$ for $\ell \geq 8$. The gap $\Delta(\ell)$ 
between the achievable combinatorial dimension $k_{\ell}$ of ${\cal C}[\ell]$ and the information-theoretic limit, i.e.,
\bean
\Delta (\ell) = \left\lfloor \log_2 {2^{\ell} \choose \ell } \right\rfloor - k_{\ell}
\eean
is bounded by $\Delta(\ell) \leq \bigl(1+\tfrac{1}{2\ln2}\bigr) \ell - \tfrac{3}{2}\log_2 \ell - \tfrac{\ln(2\pi/e)}{2\ln 2}$ by \eqref{eq:ubi1} and Prop.~\ref{prop:klbound}. We observe that $\Delta(\ell)$ grows strictly slower than the quadratic growth of both $k_{\ell}$ and the upper bound with respect to $\ell$. (See Fig~\ref{fig:kl}.)

    \begin{figure}
    \begin{center}
        \includegraphics[width=2.7in]{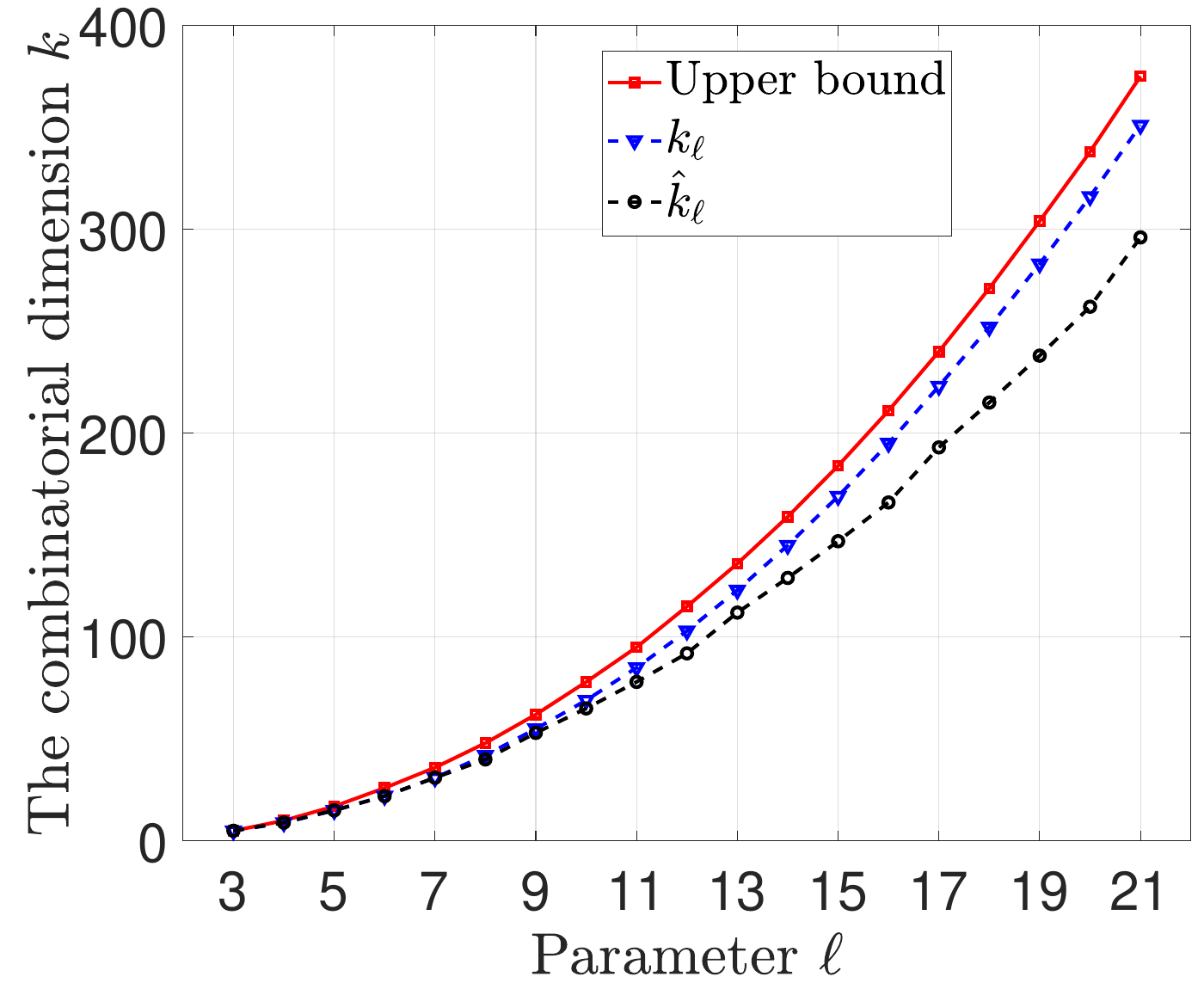}
        \caption{\centering Comparison of $k_{\ell}$ and $\hat{k}_{\ell}$ against the upper bound as $\ell$ varies.\label{fig:kl}}
    \end{center}
    \end{figure}


\subsection{Encoding and Decoding Complexities \label{subsec:complex}}

The encoding algorithm (Algorithm~\ref{alg:g1}) clearly has linear time-complexity in the input size. Both the decoding algorithms (Algorithm~\ref{alg:g2} and Algorithm~\ref{alg:g4}) involve three important steps: (a) parsing the input of length $n=2^\ell$ to identify the gap vector of length $\ell$, (b) parsing the gap vector to identify the starting point, and finally (c) converting $\ell$ gap values to their binary representation. Each step has time complexity $O(n)$, $O(\ell)=O(\log n)$ and $O(\ell^2)=O(\log^2 n)$ respectively. Except for the first round of parsing the input to obtain the gaps that is clearly linear in input size $n$, the remaining part has a poly-logarithmic time-complexity in input size. Whereas the Algorithm~\ref{alg:g2} computes only the maximum value among the gap vector, the Algorithm~\ref{alg:g4} needs to compute all gaps above a particular threshold. Therefore, despite that both have the same order of complexity, Algorithm~\ref{alg:g4} has larger time-complexity if we consider constants.

The encoding/decoding algorithms of most of the constant weight codes involve computation of binomial coefficients. One way to circumvent this problem is to store these coefficients as lookup tables, but in that case it consumes large space complexity. For example, a classic encoding (unranking) algorithm based on combinadics \cite{GenP21} requires storage of around $w{n \choose w}$ binomial coefficients. Our algorithms fully eliminate the need to compute binomial coefficients.

\section{Derived Codes \label{sec:derived}}

In this section, we derive new codes from the codes described in Sec.~\ref{sec:main} and Sec.~\ref{sec:second} by suitable  transformations that help to enlarge the parameter space. In certain range of parameters, they also achieve the information-theoretic upper bound on its size. Though we describe these new codes taking ${\cal C}[\ell]$ as the base code, similar transformations are applicable for ${\cal C}[s_{\ell}]$ and ${\cal \hat{C}}[\ell]$ as well.

\subsection{Enlarging the Range of Weight \label{subsec:wtrange}}

We present two different ways to enlarge the range of weight parameter. 

\subsubsection{${\cal C}_{t}[\ell]$: By Modifying the Sequence \label{subsec:clt}} 
Let $\ell \geq 3$ and $t$ be positive integers such that $\log_2 t < \ell-1$. Then we define a sequence $f_{\ell}^{(t)}$ of length $t$ as follows. If $t$ is not a power of $2$,
\bean
f_{\ell}^{(t)}(i) & = & \left\{ \begin{array}{ll}
     \ell - \lceil \log_2 t \rceil, & i=1,2,\ldots,t-\mu  \\
     \ell - \lfloor \log_2 t \rfloor, & i=t-\mu+1,t-\mu+2,\ldots, t-1 \\
     \ell, & i=t
\end{array} \right.
\eean
where $\mu_t = 2^{\lceil \log_2 t \rceil} - t$. If $t$ is a power of $2$, then
\bean
f_{\ell}^{(t)}(i) & = & \left\{ \begin{array}{ll}
     \ell - \log_2 t - 1, & i=1\\
     \ell - \log_2 t, & i=2,3,\ldots, t-1 \\
     \ell, & i=t
\end{array} \right.
\eean 
The construction of ${\cal C}[\ell]$ and related theorems developed in Sec.~\ref{sec:main} holds true even with respect to $f_{\ell}^{(t)}$ if we suitably modify the encoding and decoding algorithms so as to take into account the change in length of the sequence. To be precise, the necessary changes are the following:
\ben
\item The algorithm (Alg.~\ref{alg:g1}) will be invoked with $f_{\ell}^{(t)}$ as the auxiliary input. The input ${\bf x}$ will be split as concatenation of $t$ binary strings ${\bf x}={\bf x}_t\Vert {\bf x}_{t-1}\Vert \cdots \Vert {\bf x}_1$ where $|{\bf x}_i|=f_{\ell}^{(t)}(i)$. Furthermore, the loop in \textsl{Line} $4$ will have $t$ iterations. 
\item In similar lines, the decoding algorithm Alg.~\ref{alg:g2} will be invoked with $f_{\ell}^{(t)}$ as the second input. The algorithm will identify $t$ locations of $1$'s in the input ${\bf c}$ at \textsl{Line} $1$ and correspondingly the gap vector $\gaps$ will have $t$ entries. The loop at \textsl{Line} $5$ will have $t-1$ iterations. The \textsc{FindAnchor} procedure will be modified to take a $t$-length vector as input. The computation of {\sf gaps\_allone} will be modified to include $f_{\ell}^{(t)}(i), i=t-1,t-2,\ldots, 1$ as its tail end.  
\een 
It is straightforward to see both the conditions of anchor-decodability can be translated for $f_{\ell}^{(t)}$ since
\bean \label{eq:udect}
2^{\ell} - \sum_{i=1}^{t-1} 2^{f_{\ell}^{(t)}(i)} & \geq & 2^{f_{\ell}^{(t)}(t-1)} 
\eean 
and the sequence $(2^{\ell} - 1 - \sum_{i=1}^{t-1} 2^{f_{\ell}^{(t)}(i)}) \Vert ( 2^{f_{\ell}^{(t)}(i)}-1, i = t-1,\ldots, 1)$ is distinguishable from its cyclic shifts. For this reason, it turns out that the output of the encoder will always lead to a vector of weight $t$ and furthermore, the decoding algorithm with the above modifications will always be correct. Thus we have a new code ${\cal C}_t[\ell]$ with parameters
\bea \label{eq:clt}
n = 2^{\ell}, \ k & = & \sum_i f_{\ell}^{(t)}(i), \ \ w = t.
\eea 
It can be checked that ${\cal C}_2[\ell]$ has $k \ = \ 2\ell-2 \ = \ \lfloor \log_2 A(2^{\ell},2,2) \rfloor$ and therefore the code ${\cal C}_2[\ell]$ is optimal for every $\ell \geq 3$.

\subsubsection{${\cal D}_t[\ell]$: By Shortening the Message \label{subsec:dlt}}

Let $\ell \geq 3$ and $t < \ell$ be positive integers. Clearly, the encoding and decoding of ${\cal C}[\ell]$ work correct even if the message vector ${\bf x}={\bf x}_\ell \Vert {\bf x}_{\ell -1}\Vert \cdots \Vert {\bf x}_1$ is shortened by setting ${\bf x}_1={\bf 0}, {\bf x}_2={\bf 0}, \ldots, {\bf x}_{\ell-t}={\bf 0}$. This simple observation leads to deriving a new constant weight code with weight $w=t$ with suitable modifications in Alg.~\ref{alg:g1} and Alg.~\ref{alg:g2}. The necessary modifications are the following.
\ben 
\item Set the last $(\ell-t)$ blocks ${\bf x}_{\ell-t}, {\bf x}_{\ell-t-1}, \ldots, {\bf x}_{1}$ to all-zero vectors. Reset those bits to $0$ that are set to $1$ in the last $\ell-t$ iterations of the loop (corresponding to ${\bf x}_{\ell-t}, {\bf x}_{\ell-t-1}, \ldots, {\bf x}_{1}$) in the encoding algorithm.
\item In the decoding algorithm, identify $t$ locations of $1$'s in the input ${\bf c}$ at \textsl{Line} $1$ and correspondingly the gap vector $\gaps$ will have $t$ entries. The loop at \textsl{Line} $5$ will have $t-1$ iterations. The \textsc{FindAnchor} procedure will be modified to take a $t$-length vector as input. Compute the {\sf gaps\_allone} vector as ${\sf gaps\_allone} = (2^{\ell} - \sum_{i=1}^{t-1} 2^{s_{\ell}(i)}) \Vert (s_{\ell}(i), i=t-1,t-2,\ldots 1)$ as a vector of length $t$.  
\een 

It is clear that the output of the modified encoder will always be a vector of weight $t$. The new decoding algorithm will be correct for the following reason. Suppose the encoding is carried out by Alg.~\ref{alg:g1} without any modifications mentioned above. Since ${\bf x}_i={\bf 0}$ for 1 $\leq i \leq \ell-t$, there will be a run of $\ell-t$ consecutive $1$'s in the output of the encoder that appears to the left (cyclically) of the gap $\gaps[{\sf anchor\_index}]$. In the modified encoder, these $1$'s are flipped to zero, and therefore $\gaps[{\sf anchor\_index}]$ is increased by $\ell-t$ whereas all the remaining gaps hold on to the same values as that of the output provided by Alg.~\ref{alg:g1}. Thus the anchor-decodability criterion is not violated and therefore the modified decoding algorithm must be correct. 

The resultant code obtained by the modified encoder is denoted by ${\cal D}_t[\ell]$ and has parameters given by:
\bea  \label{eq:dlt}
n=2^\ell, \ \ k = k_{\ell} - \sum_{i=1}^{\ell-t} f_{\ell}(i), \ \ w = t .
\eea 
It is easy to check that ${\cal D}_2[\ell]$ is exactly same as the optimal code ${\cal C}_2[\ell]$.

%
\subsection{Enlarging the Range of Blocklength \label{subsec:short_blt}}
Let $\ell \geq 3$ and let $t < f_{\ell}(1)$ be positive integers. Unlike the construction of ${\cal D}_2[\ell]$, it is possible to shorten the message vector ${\bf x}={\bf x}_\ell \Vert {\bf x}_{\ell -1}\Vert \cdots \Vert {\bf x}_1$ by setting first (most significant) $t$ bits of ${\bf x}_{\ell}$ and the last (least significant) $t$ bits of ${\bf x}_{1}$ as zero before passing it to the encoding algorithm Alg.~\ref{alg:g1}. This leads to a constant weight code ${\cal B}_t[\ell]$ with smaller size, reduced blocklength but with the same weight $\ell$, provided that the encoding algorithm is adjusted with suitable modifications. The code ${\cal B}_t[\ell]$ has parameters
\bea \label{eq:blt}
n = 2^{\ell} - 2^t + 1, \ \ k \ = \ k_{\ell} - 2t, \ \ w \ = \ \ell.
\eea
The modified encoding algorithm is presented in Algorithm~\ref{alg:g6}. In spite of the reduction in blocklength, the weight still remains as $\ell$ as shown in Lemma~\ref{lem:bltwt}.
\begingroup
\begin{algorithm}
	\caption{\textsc{EncodeB} \newline {\bf Input}: ${\bf x} \in \{0,1\}^{k_{\ell}-t}$, $f_{\ell}$ \newline {\bf Output}: ${\bf c} \in {\cal B}_t[\ell]$ \label{alg:g6}}
	\DontPrintSemicolon
	Partition ${\bf x}$ as ${\bf x}_{\ell}\Vert {\bf x}_{\ell-1} \Vert \cdots \Vert {\bf x}_{1}$ such that $|{\bf x}_{\ell}|=\ell-t$, $|{\bf x}_{i}|=f_{\ell}(i), 2 \leq i \leq \ell-1$ and $|{\bf x}_{1}|=f_{\ell}(1)-t$.

    Initialize array ${\bf c}=0^{2^{\ell}}$ 
    
    ${\sf pos} \leftarrow 2^t\text{dec}({\bf x}_{\ell})$

    $c[{\sf pos}] \xleftarrow{} 1$
    
    \For{$j = \ell-1,\ldots,1$}    
    { 
        ${\sf pos} \xleftarrow{} {\sf pos} + 1 + \text{dec}({\bf x}_j) \mod n$

        $c[{\sf pos}] \xleftarrow{} 1$
    }
 
    ${\bf c} \leftarrow \Bar{\bf c}[{\sf pos}+1,2^t-1]$
\end{algorithm}
\endgroup
\blem \label{lem:bltwt} For every output ${\bf c}$ of Algorithm~\ref{alg:g6}, $w_H({\bf c}) = \ell$.
\elem 
\bpf Consider ${\bf c}$ in Algorithm~\ref{alg:g6} before \textsl{Line} $8$ is executed. Recall the proof of Lemma~\ref{lem:wt} and in particular \eqref{eq:p} that estimates the maximum cumulative increment $p$ in the variable ${\sf pos}$. Applying that to the context of Algorithm~\ref{alg:g6}, we observe that $p$ by the end of $(\ell-1)$ iterations of the loop at \textsl{Line} $5$ satisfies
\bea \label{eq:pb}
p & \leq & 2^{\ell} - 2^{f_{\ell}(\ell-1)} - 2^t .
\eea 
So the truncation of ${\bf c}$ by $2^t-1$ effected by the execution of \textsl{Line} $8$ does not lead to removal of a bit with value $1$. Therefore the output ${\bf c}$ has Hamming weight $\ell$.
\epf 
\begingroup
\begin{algorithm}
	\caption{\textsc{DecodeB}  \newline {\bf Input}: ${\bf c} \in {\cal B}_t[\ell], f_{\ell} $ \newline {\bf Output}: ${\bf x} \in \{0,1\}^{k_{\ell}-2t}$ \label{alg:g7}}
	\DontPrintSemicolon
	Find $0 \leq j[0] < j[1] < \cdots < j[\ell -1 ] < 2^n$ such that $c[j[i]] = 1$ for every $i=0,1,\ldots, \ell-1$. 
    
    $\gaps[m] = (j[m] - j[(m-1) \mod \ell] - 1)\mod n$ for $m=0,1,\ldots \ell-1$

    ${\sf anchor\_index} = \textsc{FindAnchorB}(\ell,t,{\bf g})$

    Initialize binary vector ${\bf x}$ such that $|{\bf x}|=\ell-r$ and $\text{dec}({\bf x})=\lceil j[{\sf anchor\_index}]/2^r\rceil$
    
    \For{$i = 1,2,\ldots, \ell-1$}    
    {
		$g \leftarrow \gaps[({\sf anchor\_index}+i) \mod \ell]$
		
		Represent $g$ as binary string ${\bf x}_{i}$ of length $\hat{f}_{\ell, r}(\ell-i)$

		${\bf x} \leftarrow {\bf x}\Vert {\bf x}_{i}$

    }
\end{algorithm}
\endgroup
\begingroup
\begin{algorithm}
	\caption{\textsc{FindAnchorB}  \newline {\bf Input}: ${\bf g} \in \mathbb{Z}_n^{\ell}, t, f_{\ell}$ \newline {\bf Output}: ${\sf anchor\_index} \in [0 \ \ell-1]$ \label{alg:g8}}
	\DontPrintSemicolon
    ${\sf gaps\_allone} \leftarrow (2^{\ell}-1-\sum_i 2^{f_{\ell}(i)} + (1-2^{-t})2^{f_{\ell}(1)}) \Vert (2^{f_{\ell}(i)}-1,i=\ell-1,\ldots, 2) \Vert (2^{f_{\ell}(1)-t}-1)$
    
    \If{ $\exists n_0 \in \mathbb{Z}_{\ell}$ \emph{such that} ${\sf gaps\_allone} = {\sf cshift}({\bf g},n_0)$ }
    {
        ${\sf anchor\_index} \leftarrow n_0$
        
    } \Else
    {
        ${\sf anchor\_index} = \arg\max_m \{{\bf g}[m] \mid m = 0,1,\ldots, \ell-1\}$
    }
\end{algorithm}
\endgroup
The decoding algorithm (presented in Algorithm~\ref{alg:g7}) is exactly in line with Algorithm~\ref{alg:g2}, but with necessary modifications to take care of the reduced length. The correctness of the decoder is established in Lemma~\ref{lem:decodeblt}. 
    
\blem \label{lem:decodeblt} For every output ${\bf c}$ of Algorithm~\ref{alg:g6}, ${\bf c}$ is correctly decoded by Algorithm~\ref{alg:g7}. 
\elem 
\bpf The encoding algorithm of ${\cal B}_t[\ell]$ differs from Alg.~\ref{alg:g1} in two aspects. First, the location $j[{\sf anchor\_index}]$ (recall the definition in \eqref{eq:init}) is the product of $2^t$ and ${\sf dec}({\bf x}_{\ell})$. Second, $(2^t-1)$ bits are deleted by \textsl{Line} $8$ of the encoding algorithm, thus reducing the length of the codeword to $2^\ell - 2^t+1$. By Lemma~\ref{lem:bltwt}, all these deleted bits are zeros. Therefore, the deletion only affects ${\bf g}[{\sf anchor\_index}]$ that do not carry any information regarding ${\bf x}_{\ell -1}, {\bf x}_{\ell -2}\ldots, {\bf x}_{1}$. As a consequence, if $j[{\sf anchor\_index}]$ is identified correctly, then ${\bf x}_{\ell -1}, {\bf x}_{\ell -2}\ldots, {\bf x}_{1}$ will be decoded correctly. 

Because of the relative decrease in ${\bf g}[{\sf anchor\_index}]$ caused by deletion of bits, the value of $j[{\sf anchor\_index}]$ can be less than the corresponding value in Alg.~\ref{alg:g2} by an amount that can at most be $2^t-1$. By \eqref{eq:pb}, in spite of a reduction by $2^t-1$ on its value, ${\bf g}[{\sf anchor\_index}]$ and hence ${\sf anchor\_index}$ will be correctly identified by $\textsc{FindAnchorB}$ procedure. However, as noted above, the value of $j[{\sf anchor\_index}]$ can be less by an amount that can at most be $2^t-1$. On the other hand, by \textsl{Line} $3$ of the encoder (Alg.~\ref{alg:g6}), the ${\sf anchor\_index}$ is obtained after multiplying $\text{dec}({\bf x}_{\ell})$ by $2^t$. Therefore, $\lceil j[{\sf anchor\_index}]/2^t \rceil$ recovers the value of ${\sf dec}({\bf x}_{\ell})$ correctly despite the shift in $j[{\sf anchor\_index}]$ by at most $2^t-1$.  Thus ${\bf x}_{\ell}$ is decoded correctly establishing that Algorithm~\ref{alg:g7} is correct.
\epf 

\section{Conclusion and Future Work\label{sec:con}}
Binary constant weight codes find extensive applications in many engineering problems such as source compression~\cite{DaiZ03}, data storage~\cite{Kur11}, design of spherical codes for communication over Gaussian channels~\cite{EriZ01}, optical communication~\cite{ChuK90}, spread-spectrum communication~\cite{DinFFJM09}, and cryptography~\cite{FinGS07}. Therefore the design of such codes with low-complexity encoding and decoding algorithms becomes quite relevant in practice. In this paper, we present several families of binary constant weight codes supporting a wide range of parameters while permitting linear encoding complexity and poly-logarithmic (discounting the linear time spent on parsing the input) decoding complexity. The present work opens up new directions for exploration such as: (a) enlarging the codebook further by controlled compromise on complexity, (b) achieving larger minimum distance by reducing the codebook size, and (c) study of correlation properties of the code. 

\section{Declarations}

\bit
\item This work is supported by the Australian Research Council through the Discovery Project under Grant DP200100731.  
\item The authors have no competing interests to declare that are relevant to the content of this article.
\item This article does not have any associated data.
\eit 
\bibliography{map}


\begin{thebibliography}{29}
\ifx \bisbn   \undefined \def \bisbn  #1{ISBN #1}\fi
\ifx \binits  \undefined \def \binits#1{#1}\fi
\ifx \bauthor  \undefined \def \bauthor#1{#1}\fi
\ifx \batitle  \undefined \def \batitle#1{#1}\fi
\ifx \bjtitle  \undefined \def \bjtitle#1{#1}\fi
\ifx \bvolume  \undefined \def \bvolume#1{\textbf{#1}}\fi
\ifx \byear  \undefined \def \byear#1{#1}\fi
\ifx \bissue  \undefined \def \bissue#1{#1}\fi
\ifx \bfpage  \undefined \def \bfpage#1{#1}\fi
\ifx \blpage  \undefined \def \blpage #1{#1}\fi
\ifx \burl  \undefined \def \burl#1{\textsf{#1}}\fi
\ifx \doiurl  \undefined \def \doiurl#1{\url{https://doi.org/#1}}\fi
\ifx \betal  \undefined \def \betal{\textit{et al.}}\fi
\ifx \binstitute  \undefined \def \binstitute#1{#1}\fi
\ifx \binstitutionaled  \undefined \def \binstitutionaled#1{#1}\fi
\ifx \bctitle  \undefined \def \bctitle#1{#1}\fi
\ifx \beditor  \undefined \def \beditor#1{#1}\fi
\ifx \bpublisher  \undefined \def \bpublisher#1{#1}\fi
\ifx \bbtitle  \undefined \def \bbtitle#1{#1}\fi
\ifx \bedition  \undefined \def \bedition#1{#1}\fi
\ifx \bseriesno  \undefined \def \bseriesno#1{#1}\fi
\ifx \blocation  \undefined \def \blocation#1{#1}\fi
\ifx \bsertitle  \undefined \def \bsertitle#1{#1}\fi
\ifx \bsnm \undefined \def \bsnm#1{#1}\fi
\ifx \bsuffix \undefined \def \bsuffix#1{#1}\fi
\ifx \bparticle \undefined \def \bparticle#1{#1}\fi
\ifx \barticle \undefined \def \barticle#1{#1}\fi
\bibcommenthead
\ifx \bconfdate \undefined \def \bconfdate #1{#1}\fi
\ifx \botherref \undefined \def \botherref #1{#1}\fi
\ifx \url \undefined \def \url#1{\textsf{#1}}\fi
\ifx \bchapter \undefined \def \bchapter#1{#1}\fi
\ifx \bbook \undefined \def \bbook#1{#1}\fi
\ifx \bcomment \undefined \def \bcomment#1{#1}\fi
\ifx \oauthor \undefined \def \oauthor#1{#1}\fi
\ifx \citeauthoryear \undefined \def \citeauthoryear#1{#1}\fi
\ifx \endbibitem  \undefined \def \endbibitem {}\fi
\ifx \bconflocation  \undefined \def \bconflocation#1{#1}\fi
\ifx \arxivurl  \undefined \def \arxivurl#1{\textsf{#1}}\fi
\csname PreBibitemsHook\endcsname

\bibitem[\protect\citeauthoryear{Johnson}{1962}]{Joh62}
\begin{barticle}
\bauthor{\bsnm{Johnson}, \binits{S.M.}}:
\batitle{A new upper bound for error-correcting codes}.
\bjtitle{{IRE} Trans. Inf. Theory}
\bvolume{8}(\bissue{3}),
\bfpage{203}--\blpage{207}
(\byear{1962})
\end{barticle}
\endbibitem

\bibitem[\protect\citeauthoryear{Graham and Sloane}{1980}]{GraS80}
\begin{barticle}
\bauthor{\bsnm{Graham}, \binits{R.}},
\bauthor{\bsnm{Sloane}, \binits{N.}}:
\batitle{Lower bounds for constant weight codes}.
\bjtitle{IEEE Transactions on Information Theory}
\bvolume{26}(\bissue{1}),
\bfpage{37}--\blpage{43}
(\byear{1980})
\doiurl{10.1109/TIT.1980.1056141}
\end{barticle}
\endbibitem

\bibitem[\protect\citeauthoryear{Brouwer et~al.}{1990}]{BroSSS90}
\begin{barticle}
\bauthor{\bsnm{Brouwer}, \binits{A.E.}},
\bauthor{\bsnm{Shearer}, \binits{J.B.}},
\bauthor{\bsnm{Sloane}, \binits{N.J.A.}},
\bauthor{\bsnm{Smith}, \binits{W.D.}}:
\batitle{A new table of constant weight codes}.
\bjtitle{IEEE Transactions on Information Theory}
\bvolume{36}(\bissue{6}),
\bfpage{1334}--\blpage{1380}
(\byear{1990})
\doiurl{10.1109/18.59932}
\end{barticle}
\endbibitem

\bibitem[\protect\citeauthoryear{Agrell et~al.}{2000}]{AgrVZ00}
\begin{barticle}
\bauthor{\bsnm{Agrell}, \binits{E.}},
\bauthor{\bsnm{Vardy}, \binits{A.}},
\bauthor{\bsnm{Zeger}, \binits{K.}}:
\batitle{Upper bounds for constant-weight codes}.
\bjtitle{IEEE Transactions on Information Theory}
\bvolume{46}(\bissue{7}),
\bfpage{2373}--\blpage{2395}
(\byear{2000})
\doiurl{10.1109/18.887851}
\end{barticle}
\endbibitem

\bibitem[\protect\citeauthoryear{Brouwer}{2023}]{aeb}
\begin{botherref}
\oauthor{\bsnm{Brouwer}, \binits{A.}}:
Bounds for binary constant weight codes.
\url{https://www.win.tue.nl/%7eaeb/codes/Andw.html}.
[Online; accessed 23-Oct-2023]
(2023)
\end{botherref}
\endbibitem

\bibitem[\protect\citeauthoryear{Moreno et~al.}{1995}]{MorZKZ95}
\begin{barticle}
\bauthor{\bsnm{Moreno}, \binits{O.}},
\bauthor{\bsnm{Zhang}, \binits{Z.}},
\bauthor{\bsnm{Kumar}, \binits{P.V.}},
\bauthor{\bsnm{Zinoviev}, \binits{V.A.}}:
\batitle{New constructions of optimal cyclically permutable constant weight
  codes}.
\bjtitle{IEEE Transactions on Information Theory}
\bvolume{41}(\bissue{2}),
\bfpage{448}--\blpage{455}
(\byear{1995})
\doiurl{10.1109/18.370146}
\end{barticle}
\endbibitem

\bibitem[\protect\citeauthoryear{Bitan and Etzion}{1995}]{BitE95}
\begin{barticle}
\bauthor{\bsnm{Bitan}, \binits{S.}},
\bauthor{\bsnm{Etzion}, \binits{T.}}:
\batitle{Constructions for optimal constant weight cyclically permutable codes
  and difference families}.
\bjtitle{IEEE Transactions on Information Theory}
\bvolume{41}(\bissue{1}),
\bfpage{77}--\blpage{87}
(\byear{1995})
\doiurl{10.1109/18.370117}
\end{barticle}
\endbibitem

\bibitem[\protect\citeauthoryear{Nordio and Viterbo}{2003}]{NorV03}
\begin{bchapter}
\bauthor{\bsnm{Nordio}, \binits{A.}},
\bauthor{\bsnm{Viterbo}, \binits{E.}}:
\bctitle{Permutation modulation for fading channels}.
In: \bbtitle{10th International Conference on Telecommunications, 2003. ICT
  2003.},
vol. \bseriesno{2},
pp. \bfpage{1177}--\blpage{11832}
(\byear{2003}).
\doiurl{10.1109/ICTEL.2003.1191603}
\end{bchapter}
\endbibitem

\bibitem[\protect\citeauthoryear{MacWilliams and Sloane}{1977}]{MacSloane}
\begin{bbook}
\bauthor{\bsnm{MacWilliams}, \binits{F.J.}},
\bauthor{\bsnm{Sloane}, \binits{N.J.A.}}:
\bbtitle{The Theory of Error-correcting Codes}.
\bsertitle{Mathematical Library}.
\bpublisher{North-Holland Publishing Company},
\blocation{New York}
(\byear{1977})
\end{bbook}
\endbibitem

\bibitem[\protect\citeauthoryear{Schalkwijk}{1972}]{Sch72}
\begin{barticle}
\bauthor{\bsnm{Schalkwijk}, \binits{J.}}:
\batitle{An algorithm for source coding}.
\bjtitle{IEEE Transactions on Information Theory}
\bvolume{18}(\bissue{3}),
\bfpage{395}--\blpage{399}
(\byear{1972})
\end{barticle}
\endbibitem

\bibitem[\protect\citeauthoryear{Cover}{1973}]{Cov73}
\begin{barticle}
\bauthor{\bsnm{Cover}, \binits{T.}}:
\batitle{Enumerative source encoding}.
\bjtitle{IEEE Transactions on Information Theory}
\bvolume{19}(\bissue{1}),
\bfpage{73}--\blpage{77}
(\byear{1973})
\end{barticle}
\endbibitem

\bibitem[\protect\citeauthoryear{Lehmer}{1960}]{Leh60}
\begin{bchapter}
\bauthor{\bsnm{Lehmer}, \binits{D.H.}}:
\bctitle{Teaching combinatorial tricks to a computer}.
In: \bbtitle{Proc. {S}ympos. {A}ppl. {M}ath., {V}ol. 10},
pp. \bfpage{179}--\blpage{193}.
\bpublisher{Amer. Math. Soc., Providence, RI},
\blocation{New York}
(\byear{1960})
\end{bchapter}
\endbibitem

\bibitem[\protect\citeauthoryear{Pascal}{1887}]{Pas87}
\begin{barticle}
\bauthor{\bsnm{Pascal}, \binits{E.}}:
\batitle{Sopra una formula numerica}.
\bjtitle{Gi Di Mat}
\bvolume{25},
\bfpage{45}--\blpage{49}
(\byear{1887})
\end{barticle}
\endbibitem

\bibitem[\protect\citeauthoryear{Knott}{1974}]{Kno74}
\begin{barticle}
\bauthor{\bsnm{Knott}, \binits{G.D.}}:
\batitle{A numbering systems for combinations}.
\bjtitle{Commun. {ACM}}
\bvolume{17}(\bissue{1}),
\bfpage{45}--\blpage{46}
(\byear{1974})
\end{barticle}
\endbibitem

\bibitem[\protect\citeauthoryear{Er}{1985}]{Er85}
\begin{barticle}
\bauthor{\bsnm{Er}, \binits{M.C.}}:
\batitle{Lexicographic ordering, ranking and unranking of combinations}.
\bjtitle{Int. J. Comput. Math.}
\bvolume{17}(\bissue{1}),
\bfpage{277}--\blpage{283}
(\byear{1985})
\end{barticle}
\endbibitem

\bibitem[\protect\citeauthoryear{Kokosinski}{1995}]{Kok95}
\begin{bchapter}
\bauthor{\bsnm{Kokosinski}, \binits{Z.}}:
\bctitle{Algorithms for unranking combinations and their applications}.
In: \beditor{\bsnm{Hamza}, \binits{M.H.}} (ed.)
\bbtitle{Proceedings of the Seventh {IASTED/ISMM} International Conference on
  Parallel and Distributed Computing and Systems, Washington, D.C., USA,
  October 19-21, 1995},
pp. \bfpage{216}--\blpage{224}
(\byear{1995})
\end{bchapter}
\endbibitem

\bibitem[\protect\citeauthoryear{Ruskey and Williams}{2009}]{RusW09}
\begin{barticle}
\bauthor{\bsnm{Ruskey}, \binits{F.}},
\bauthor{\bsnm{Williams}, \binits{A.}}:
\batitle{The coolest way to generate combinations}.
\bjtitle{Discret. Math.}
\bvolume{309}(\bissue{17}),
\bfpage{5305}--\blpage{5320}
(\byear{2009})
\end{barticle}
\endbibitem

\bibitem[\protect\citeauthoryear{Genitrini and P{\'{e}}pin}{2021}]{GenP21}
\begin{barticle}
\bauthor{\bsnm{Genitrini}, \binits{A.}},
\bauthor{\bsnm{P{\'{e}}pin}, \binits{M.}}:
\batitle{Lexicographic unranking of combinations revisited}.
\bjtitle{Algorithms}
\bvolume{14}(\bissue{3}),
\bfpage{97}
(\byear{2021})
\end{barticle}
\endbibitem

\bibitem[\protect\citeauthoryear{Kruchinin et~al.}{2022}]{KruSKR22}
\begin{barticle}
\bauthor{\bsnm{Kruchinin}, \binits{V.V.}},
\bauthor{\bsnm{Shablya}, \binits{Y.V.}},
\bauthor{\bsnm{Kruchinin}, \binits{D.V.}},
\bauthor{\bsnm{Rulevskiy}, \binits{V.}}:
\batitle{Unranking small combinations of a large set in co-lexicographic
  order}.
\bjtitle{Algorithms}
\bvolume{15}(\bissue{2}),
\bfpage{36}
(\byear{2022})
\end{barticle}
\endbibitem

\bibitem[\protect\citeauthoryear{Sendrier}{2005}]{Sen05}
\begin{bchapter}
\bauthor{\bsnm{Sendrier}, \binits{N.}}:
\bctitle{Encoding information into constant weight words}.
In: \bbtitle{Proceedings. International Symposium on Information Theory, 2005.
  ISIT 2005.},
pp. \bfpage{435}--\blpage{438}
(\byear{2005})
\end{bchapter}
\endbibitem

\bibitem[\protect\citeauthoryear{Slepian}{1965}]{Sle65}
\begin{barticle}
\bauthor{\bsnm{Slepian}, \binits{D.}}:
\batitle{Permutation modulation}.
\bjtitle{Proceedings of the IEEE}
\bvolume{53}(\bissue{3}),
\bfpage{228}--\blpage{236}
(\byear{1965})
\doiurl{10.1109/PROC.1965.3680}
\end{barticle}
\endbibitem

\bibitem[\protect\citeauthoryear{Gallager}{2008}]{Gallager_2008}
\begin{bbook}
\bauthor{\bsnm{Gallager}, \binits{R.G.}}:
\bbtitle{Principles of Digital Communication}.
\bpublisher{Cambridge University Press},
\blocation{New York}
(\byear{2008})
\end{bbook}
\endbibitem

\bibitem[\protect\citeauthoryear{Riordan}{1978}]{Riordan}
\begin{bbook}
\bauthor{\bsnm{Riordan}, \binits{J.}}:
\bbtitle{An Introduction to Combinatorial Analysis}.
\bsertitle{Princeton Legacy Library}.
\bpublisher{Princeton University Press},
\blocation{Princeton}
(\byear{1978})
\end{bbook}
\endbibitem

\bibitem[\protect\citeauthoryear{Dai and Zakhor}{2003}]{DaiZ03}
\begin{bchapter}
\bauthor{\bsnm{Dai}, \binits{V.}},
\bauthor{\bsnm{Zakhor}, \binits{A.}}:
\bctitle{Binary combinatorial coding}.
In: \bbtitle{Data Compression Conference, 2003. Proceedings. DCC 2003},
p. \bfpage{420}
(\byear{2003}).
\doiurl{10.1109/DCC.2003.1194039}
\end{bchapter}
\endbibitem

\bibitem[\protect\citeauthoryear{Kurmaev}{2011}]{Kur11}
\begin{barticle}
\bauthor{\bsnm{Kurmaev}, \binits{O.F.}}:
\batitle{Constant-weight and constant-charge binary run-length limited codes}.
\bjtitle{IEEE Transactions on Information Theory}
\bvolume{57}(\bissue{7}),
\bfpage{4497}--\blpage{4515}
(\byear{2011})
\doiurl{10.1109/TIT.2011.2145490}
\end{barticle}
\endbibitem

\bibitem[\protect\citeauthoryear{Ericson and Zinoviev}{2001}]{EriZ01}
\begin{bchapter}
\bauthor{\bsnm{Ericson}, \binits{T.}},
\bauthor{\bsnm{Zinoviev}, \binits{V.}}:
\bctitle{Chapter 6 - non-symmetric alphabets}.
In: \beditor{\bsnm{Ericson}, \binits{T.}},
\beditor{\bsnm{Zinoviev}, \binits{V.}} (eds.)
\bbtitle{Codes on Euclidean Spheres}.
\bsertitle{North-Holland Mathematical Library},
vol. \bseriesno{63},
pp. \bfpage{179}--\blpage{194}.
\bpublisher{Elsevier},
\blocation{New York}
(\byear{2001}).
\doiurl{10.1016/S0924-6509(01)80051-9}
\end{bchapter}
\endbibitem

\bibitem[\protect\citeauthoryear{Chung and Kumar}{1990}]{ChuK90}
\begin{barticle}
\bauthor{\bsnm{Chung}, \binits{H.}},
\bauthor{\bsnm{Kumar}, \binits{P.V.}}:
\batitle{Optical orthogonal codes - new bounds and an optimal construction}.
\bjtitle{IEEE Transactions on Information Theory}
\bvolume{36}(\bissue{4}),
\bfpage{866}--\blpage{873}
(\byear{1990})
\doiurl{10.1109/18.53748}
\end{barticle}
\endbibitem

\bibitem[\protect\citeauthoryear{Ding et~al.}{2009}]{DinFFJM09}
\begin{barticle}
\bauthor{\bsnm{Ding}, \binits{C.}},
\bauthor{\bsnm{Fuji-Hara}, \binits{R.}},
\bauthor{\bsnm{Fujiwara}, \binits{Y.}},
\bauthor{\bsnm{Jimbo}, \binits{M.}},
\bauthor{\bsnm{Mishima}, \binits{M.}}:
\batitle{Sets of frequency hopping sequences: Bounds and optimal
  constructions}.
\bjtitle{IEEE Transactions on Information Theory}
\bvolume{55}(\bissue{7}),
\bfpage{3297}--\blpage{3304}
(\byear{2009})
\doiurl{10.1109/TIT.2009.2021366}
\end{barticle}
\endbibitem

\bibitem[\protect\citeauthoryear{Finiasz et~al.}{2011}]{FinGS07}
\begin{bchapter}
\bauthor{\bsnm{Finiasz}, \binits{M.}},
\bauthor{\bsnm{Gaborit}, \binits{P.}},
\bauthor{\bsnm{Sendrier}, \binits{N.}}:
\bctitle{Improved fast syndrome based cryptographic hash functions}.
In: \bbtitle{ECRYPT Hash Workshop 2007, Proceedings},
p. \bfpage{155}
(\byear{2011})
\end{bchapter}
\endbibitem

\end{thebibliography}

\end{document}